\documentclass{amsart}

\usepackage{amsmath,amssymb,a4wide,amsthm,ifthen,hyperref,xcolor,enumerate,enumitem}
\usepackage{tabularx}
\usepackage{stmaryrd}
\usepackage{graphicx}
\usepackage[utf8]{inputenc}
\usepackage{natbib}
\usepackage{amsaddr}

\def\1{1\!{\rm l}}
\newcommand{\argmin}{\operatornamewithlimits{argmin}}
\newcommand{\argmax}{\operatornamewithlimits{argmax}}

\newcommand{\mubf}{\mbox{\mathversion{bold}{$\mu$}}}
\newcommand{\tbf}{\mbox{\mathversion{bold}{$t$}}}

\newcommand{\ybf}{\mbox{\mathversion{bold}{$y$}}}

\pdfoutput=1
\newcommand{\ackname}{Acknowledgements}
\newcommand{\datastat}{Data statement}

\title{A new segmentation method for the homogenisation of GNSS-derived IWV time-series}

\date{\today}

\author{A. Quarello}
\author{O. Bock}
\address[A1,A2]{Universit\'e de Paris, Institut de physique du globe de Paris, CNRS, IGN, F-75005 Paris, France. ENSG-Géomatique, IGN, F-77455 Marne-la-Vallée, France.}
\author{E. Lebarbier}
\address[A3]{Laboratoire Modal’X, UPL, Univ. Paris Nanterre, France. }

\begin{document}

%\title{A new segmentation method for the homogenisation of GNSS-derived IWV time-series}

%\author{Annarosa Quarello \thanks{Universit\'e de Paris, Institut de physique du globe de Paris, CNRS, IGN, F-75005 Paris, France. ENSG-Géomatique, IGN, F-77455 Marne-la-Vallée, France} \and Emilie Lebarbier \thanks{Laboratoire Modal’X, UPL, Univ. Paris Nanterre, France. e-mail:emilie.lebarbier@paraisnanterre.fr} \and Olivier Bock \thanks{Universit\'e de Paris, Institut de physique du globe de Paris, CNRS, IGN, F-75005 Paris, France. ENSG-Géomatique, IGN, F-77455 Marne-la-Vallée, France}}

\maketitle

\begin{abstract} Homogenization is an important and crucial step to improve the usage of observational data for climate analysis. This work is motivated by the analysis of long series of GNSS Integrated Water Vapour (IWV) data which have not yet been used in this context. This paper proposes a novel segmentation method that integrates a periodic bias and a heterogeneous, monthly varying, variance. The method consists in estimating first the variance using a robust estimator and then estimating the segmentation and periodic bias iteratively. This strategy allows for the use of the dynamic programming algorithm that remains the most efficient exact algorithm to estimate the change-point positions. The statistical performance of the method is assessed through numerical experiments. An application to a real data set of 120 global GNSS stations is presented. The method is implemented in the R package GNSSseg that will be available on the CRAN.
\end{abstract}

keywords: Change-point detection; Dynamic programming; Homogenization climate series; GNSS IWV series

\section {Introduction}  \label{sec:intro}

Long records of observational data are essential to monitoring climate change and understanding the underlying climate processes. However, long time series are often affected by inhomogeneties due to changes in instrumentation, in station location, in observation and processing methods, and/or in the measurement conditions around the station \citep{Jones1986}. Inhomogeneities most often take the form of abrupt changes which are detrimental to estimating trends and multi-scale climate variability \citep{Easterling1995}. Various homogenization methods have been developed for the detection and correction of such change-points in the context of climate data analysis, e.g. \citet{Peterson1998, CaussinusMestre2004, Menne2005, Szentimrey2008, Reeves2007, Costa2009, Venema2012}. In this paper, we are interested in ground-based Global Navigation Satellite System (GNSS) integrated water vapour (IWV) measurements. GNSS measurements provide among the most accurate and continuous IWV measurements, in all weather conditions, and have not yet been used much for climate analysis \citep{Bevis1992, Bock2013, VanMalderen2014, NingUncertainty2016}.

\begin{center}
\begin{figure}[ht]
\begin{tabular}{cc}
   ($a$) & ($b$) \\
\includegraphics[width=7cm,height=5cm]{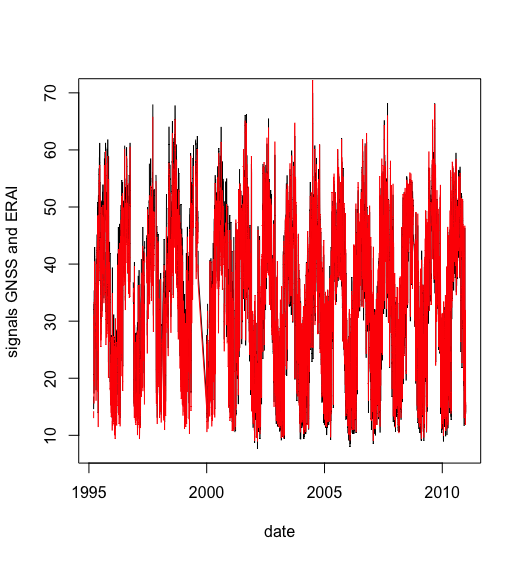} & \includegraphics[width=7cm,height=5cm]{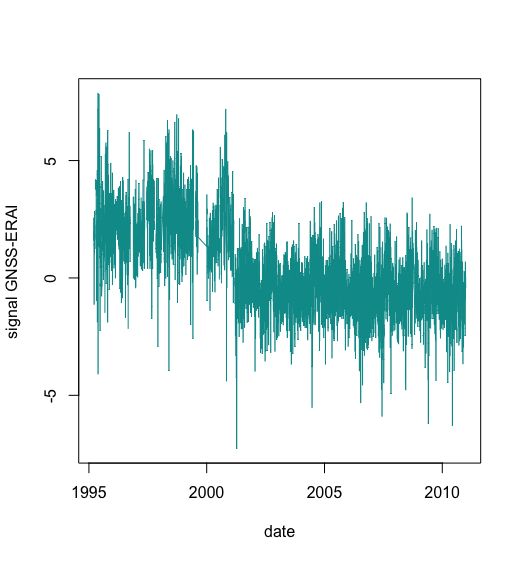} \\
($c$) & ($d$) \\
\includegraphics[width=7cm,height=5cm]{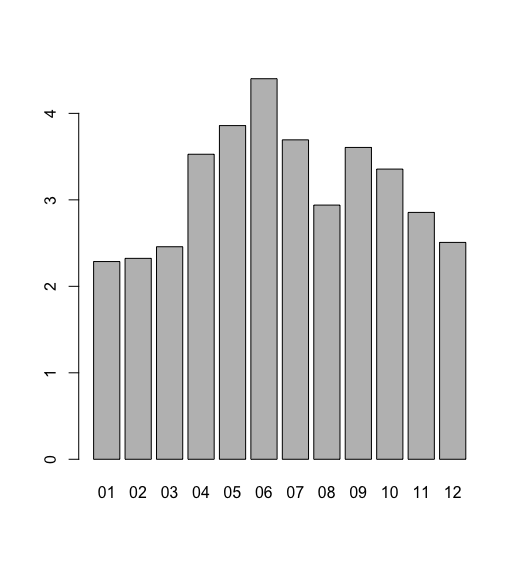}& 
\includegraphics[width=7cm,height=5cm]{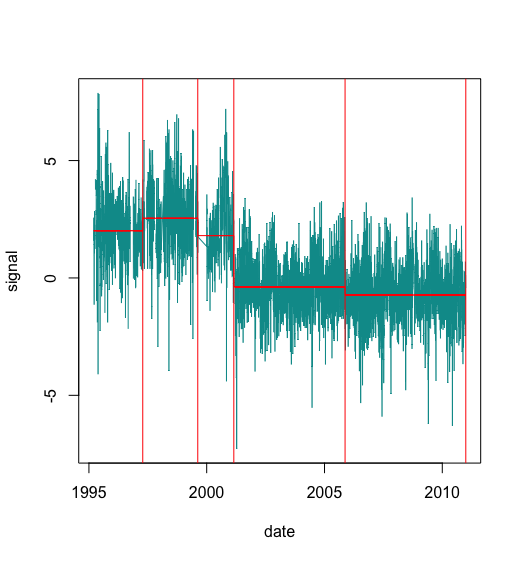}
\end{tabular}
\caption{Station CCJM: (a) GNSS (in black) and ERA-Interim (in red) IWV time series; (b) IWV difference (GPS - ERA-Interim) series; (c) estimated monthly variance; (d) obtained change-points with the method proposed by \cite{Bock2018}. 
\label{fig:ccjm}}
\end{figure}
\end{center}

In order to remove the climate signal and reveal the inhomogeneities in the GNSS measurements, it has been a common approach to compare the candidate series with a well correlated reference series. The reference series can be taken from nearby stations (i.e. observing a similar climate signal) as proposed by, e.g., \citet{CaussinusMestre2004}, \citet{Menne2005}, or \citet{Szentimrey2008}, or from climate model data \citet{ning2016,Bock2018}. Since the number of stations from the International GNSS Service (IGS) is limited to a hundred or so, the construction of references series from neighboring stations is hard. The second approach is here 
considered using the European Center for Medium Range Forecasts (ECMWF) reanalysis ERA-Interim \citep{Dee2011} as a reference. Figure \ref{fig:ccjm}(a) shows an example of daily IWV data from GNSS measurements and from the ERA-Interim (ERAI) reanalysis at station CCJM. The daily IWV data exhibit a marked seasonal variation, with values varying from 10 $kg m^{-2}$ to 60 $kg m^{-2}$ between winter and summer, as well as a strong day-to-day variability looking as superposed noise. When the ERA-Interim data are subtracted from the GNSS data, one clear jump can be seen on 24 Feb 2001 (Figure \ref{fig:ccjm}(b)). This jump coincides with a change of receiver and antenna at this station.

In a previous work, \citet{Bock2018} proposed a first segmentation method to detect abrupt changes in the mean in such data (GNSS - ERAI IWV differences). A specific feature of their method is that it accounts for a heterogeneous variance that is assumed to vary on a monthly basis. Indeed, as can be seen in Figure \ref{fig:ccjm}(c), the GNSS - ERAI IWV differences show a seasonal variation with an increased variability in summer. Thus classical segmentation models with homogeneous or segment-specific variance are not adapted. The result of their model is given in Figure \ref{fig:ccjm}(d). The previously mentioned jump is well detected. However, as already mentioned in \cite{Bock2018}, despite the ERAI data are subtracted, it can happen that not all the climate signal is removed due to representativeness differences between the reanalysis and the GNSS observations \citep{Bock2019}. This residual signal exhibits a strong seasonal variation which can lead to wrong, misplaced, or missing change-points. 

This paper described an improved method which accounts for seasonal variation in the signal by adding a functional term to earlier model used by \citet{Bock2018}. To infer the parameters of this enhanced model, a (penalized)-maximum likelihood procedure is used again. In this framework, it is well known that segmentation methods have to deal with two problems: (i) an inherent algorithmic complexity for estimating the change-point locations and (ii) an appropriate choice of the penalty term which controls the number of change-points. Indeed, for problem (i), the inference of the discrete change-points requires to search over the whole segmentation space that is huge. Such a search is prohibitive in terms of computational time when performed in a naive way. The Dynamic Programming (DP) algorithm \citep{Auger1989} and its recent pruned versions \citep{Killick2012, Rigaill2015, Maidstone2017}, are the only algorithms that retrieve the exact solution in a fast way. However, a necessary condition for using DP is that the quantity to be optimized is additive with respect to the segments \citep{BaiPerron2003, CaussinusMestre2004, Picard2005}. Here, with the presence of the monthly variance and the functional part, the condition is not verified. To circumvent this, \citet{LiLund2012} and \citet{Lu2010} proposed to use a genetic algorithm. However, this algorithm leads to a suboptimal solution. 

Our objective here is to keep the interest of the exact DP algorithm as possible. To be enable us to use it in the inference procedure, we propose to (1) estimate first the variance using a robust (to the change-points) estimator as in \citet{Bock2018} and (2) treat sequentially the estimation of the segmentation parameters and the functional as in \citet{Bertin2017}. For the choice of the number of segments, different penalties have been proposed in the literature (see \citet{Lebarbier2005, Lavielle2005, Zhang2007, Lu2010, CaussinusMestre2004}). Here we propose to use some of them. 

The article is organized as follows. Section \ref{sec:method} presents the model and the inference procedure. A simulation study is performed in Section \ref{sec:sim} to evaluate the performance of the method. In Section \ref{sec:real} the method is applied on real data from a set 120 global GNSS stations. Section \ref{sec:conclusion} discusses the results and concludes.

\section{Model and inference}\label{sec:method}

\subsection{Model}
We consider the model proposed by \citet{Bock2018} in which we add a functional part in order to take into account the periodic bias. Let  $\textbf{y}=\{y_t\}_{1,\ldots,n}$ be the observed series with length $n$ that is modeled by a Gaussian independent random process $\textbf{Y}=\{Y_t\}_{t=1,\ldots,n}$ such that
\begin{itemize}
\item the mean of $\textbf{Y}$ is composed of two terms:
\begin{itemize}
    \item[$\star$] a piecewise constant function equals to $\mu_k$ on the interval $I_k^{\text{mean}}= \llbracket t_{k-1}+1, t_{k} \rrbracket$ with length $n_k=t_{k}-t_{k-1}$ where $0=t_0<t_1<\ldots<t_{K-1}<t_K=n$. The $\{t_{k}\}_{k=1,\ldots,K-1}$ are the times of the change-points and $K$ is the number of intervals or segments,
 \item[$\star$] and a function $f$;
\end{itemize}
\item the variance of $\textbf{Y}$ is month-dependent, i.e. it is constant within the interval $I_{\text{month}}^{\text{var}}=\{t; \text{date}(t) \in \text{month}\}$ with length $n_{\text{month}}$ where $\text{date}(t)$ stands for the date at the time $t$. 
\end{itemize}
The resulting model is thus the following
\begin{equation} \label{sec:method:eq:m}
Y_t =\mu_k+f_t+ E_t,  \ \ \text{where $E_t\sim \mathcal{N}(0,\sigma_{\text{month}}^2)$ if $t \in I_{k}^{\text{mean}} \cap I^{\text{var}}_{\text{month}}$}, 
\end{equation}
%where the errors $\{E_t\}_t$ are supposed to be centered independent Gaussian with heterogeneous variance, i.e. $\{E_t\}_t \ \text{i.i.d.} \sim \mathcal{N}(0,\sigma_{\text{month}}^2)$ if $t \in I^{\text{var}}_{\text{month}}$ and 
for $k=1,\ldots,K$. %where $K$ is the number of segments or intervals $I_{k}^{\text{mean}}$.
The intervals  $\{I_k^{\text{mean}}\}_k$ are unknown contrary to the intervals $\{I_{\text{month}}^{\text{var}}\}_{\text{month}}$. The parameters to be estimated are the number of segments $K$ (or the number of change-points $K-1$), the change-points $\tbf=\{t_k\}_k$ and the distribution parameters, the means $\mubf=\{\mu_k\}_k$, the variances ${\bf{\sigma}^2}=\{\sigma^2_{\text{month}}\}_{\text{month}}$ and the function $f$.

%--------------------------------------------------------------------------
\subsection{Inference} \label{sec:method:subsec:inference}

As usual in segmentation, the inference is performed in two steps (e.g. \citet{Truong2019}): 
\begin{description}
\item[Step 1] Estimate $\tbf$, $\mubf$, ${\bf{\sigma}^2}$ and $f$, $K$ being fixed.
\item[Step 2] Choose the number of segments $K$.
\end{description}
We consider here a penalized maximum likelihood approach. The $\log$-likelihood of the model defined by \eqref{sec:method:eq:m} is 
\begin{equation} \label{eq:loglik_m}
\footnotesize
\log \ p(\ybf; K, \tbf, \mubf, {\bf{\sigma}^2},f)= - \frac{n}{2} \log{(2 \pi)} - \sum_{\text{month}} \frac{n_{\text{month}} }{2}  \log{(\sigma^2_{\text{month}})}-\frac{1}{2} \sum_{k=1}^K \sum_{\text{month}} \sum_{t \in I_k^{\text{mean}} \cap I^{\text{var}}_\text{month}} \frac {(y_t-\mu_k-f_t)^2}{\sigma^2_{\text{month}}}
\end{equation}

\subsubsection{Step 1: Inference of $\tbf$, $\mubf$, ${\bf{\sigma}^2}$ and $f$, $K$ being fixed} \label{sec:method:subsec:inference:subsub:inference_K}

The use of the DP algorithm is now classical to estimate the change-points. However, DP can be applied if and only if the quantity to be optimized is additive with respect to the segments. Here the presence of the 'global' parameters $\sigma^2_{\text{month}}$ and $f$ will link the segments and the required condition will not be satisfied. To circumvent this problem a two-step procedure is proposed: (1) we estimate the variances using a robust estimator as in \cite{Chakar2017} and \cite{Bock2018} and (2) we estimate iteratively $f$ and the segmentation parameters (i.e. the change-points and the means) using DP as in \cite{gazeaux2015joint} and \cite{Bertin2017}.

The resulting algorithm is the following: 
\begin{description}

\item[Estimation of $\sigma^2_{\text{month}}$] \citet{Bock2018} proposed a consistent estimator for the variance parameter based on the robust one proposed by \citet{Rousseeuw1993}. The key idea is to apply this robust estimator (up to a constant) on the differentiated series $y_t-y_{t-1}$. This series is centered except at the change-point positions which are treated as outliers. We again use this estimator even in the presence of the function $f$ because the latter does not have much impact on the resulting estimation (in the application, the seasonal signal is slowly varying and is almost completely cancelled in the differentiated series). The estimated variance is noted $\widehat{\sigma}_{\text{month}}^{2}$. 

\item[Estimation of $f$ and both $\tbf$ and $\mubf$ iteratively] by minimizing the minus $\log$-likelihood given in \eqref{eq:loglik_m}. At iteration $[h+1]$:
\begin{itemize}
\item[$(a)$] The estimator of $f$ results in a weighted least-square estimator with weights $1/\widehat{\sigma}_{\text{month}}^{2}$ on $\{y_t -{\mu}_k^{[h]}\}_t$. For our application and following \cite{Weatherhead1998}, we represent $f$ as a Fourier series of order $4$ accounting for annual, semi-annual, terannual, and quarterly periodicities in the signal:
\begin{equation*} 
f_t = \sum_{i=1}^{4} a_{i} \cos( w_i t) + b_{i} \sin( w_i t), \\
\end{equation*}
where $w_i= 2\pi \frac{i}{L}$ is the angular frequency of period $L/i$ and $L$ is the mean length of the year ($L=365.25$ days when time $t$ is expressed in days). The estimated function is denoted $f^{[h+1]}$. 

\item[$(b)$] The segmentation parameters are estimated based on $\{y_t -{f}^{[h+1]}_t\}_t$. We get
\begin{equation*} 
{\mu}^{[h+1]}_k = \frac{\sum_{\text{month}} \sum_{t \in I_k^{\text{mean}} \cap I^{\text{var}}_{\text{month}}} \frac{({y}_t-{f}^{[h+1]}_t)}{\widehat{\sigma}_{\text{month}}^{2}}} { \sum_{\text{month}}  \sum_{t \in I_k^{\text{mean}} \cap I^{\text{var}}_{\text{month}}} \frac{1}{\widehat{\sigma}_{\text{month}}^{2}}}, 
\end{equation*}
and
\begin{equation*} 
{\tbf}^{[h+1]} = \argmin_{\tbf \in \mathcal{M}_{K,n}} \sum_{k=1}^K \sum_{\text{month}} \sum_{t \in I_k^{\text{mean}} \cap I^{\text{var}}_{\text{month}}} \frac{({y}_t-{f}^{[h+1]}_t-{\mu}^{[h+1]}_k)^2}{\widehat{\sigma}_{\text{month}}^{2}},
\end{equation*}
where $\mathcal{M}_{K,n}=\{(t_1,\ldots,t_{K-1}) \in \mathbb{N}^{K-1}, 0=t_0<t_1<\ldots,t_{K-1}<t_K=n\}$ is the set of all the possible partitions of the grid $\llbracket 1, n \rrbracket$ in $K$ segments. This minimization is obtained using DP. 
\end{itemize}
The final estimators are denoted $\widehat{f}$, $\widehat{\tbf}$ and $\widehat{\mubf}$.

\end{description}

\subsubsection{Choice of $K$} \label{sec:method:subsec:inference:subsub:ModelSelection}
Various criteria have been theoretically developed for the choice of $K$ in segmentation with a homogeneous (known or unknown) variance. However, no criteria exist for the case with a heterogeneous variance on fixed intervals. Since in our estimation procedure the variances are estimated first, our segmentation problem can be seen as one in which the variance is known. We thus propose to use the least-squares-based criterion:
\begin{equation} 
\text{SSR}_K (\widehat{\tbf}, \widehat{\mubf}, {\widehat{\bf{\sigma}}}^2,\widehat{f})=\sum_{k=1}^K \sum_{\text{month}} \sum_{t \in \widehat{I}^{\text{mean}}_k \cap I^{\text{var}}_{\text{month}}} \frac{({y}_t-\widehat{f}_t-\widehat{\mu}_k)^2}{\widehat{\sigma}_{month}^{2}}. 
\end{equation}
Different penalties are considered and tested in this paper: 
\begin{description}
\item[Lav] proposed by \cite{Lavielle2005}: 
\begin{equation*}
\widehat{K}=\argmin_{K} \text{SSR}_K (\widehat{\tbf}, \widehat{\mubf}, {\widehat{\bf{\sigma}}}^2,\widehat{f})+\beta K,
\end{equation*}
where $\beta$ is the penalty constant chosen using an adaptive method. The method involves a threshold $S$ which is fixed to $S=0.75$, both in the simulation study and the applications, as suggested by \citet{Lavielle2005}.
\item[BM] proposed by \cite{BM2001} and \cite{Lebarbier2005} for an application in a segmentation context:
\begin{equation*} 
\widehat{K}=\argmin_{K} \text{SSR}_K (\widehat{\tbf}, \widehat{\mubf}, {\widehat{\bf{\sigma}}}^2,\widehat{f})+\alpha K \left[ 5  +2  \log \left (\frac{n}{K} \right) \right],
\end{equation*}
where the penalty constant $\alpha$ can be calibrated using the slope heuristic proposed by \cite{Arlot2009}. Two methods are proposed actually: the "dimension jump" and the "data-driven slope estimation" which are referred to as BM1 and BM2, respectively, hereafter. 
\item[mBIC] the modified version of the classical BIC criterion derived in the segmentation framework by \cite{Zhang2007}, 
\begin{equation*} 
\widehat{K}=\argmax_{K} - \frac{1}{2} \text{SSR}_K (\widehat{\tbf}, \widehat{\mubf}, {\widehat{\bf{\sigma}}}^2,\widehat{f}) -\frac{1}{2} \sum_{k=1}^K \log {(\widehat{t}_k - \widehat{t}_{k-1})} +\left (\frac{1}{2}-K \right ) \log{(n)}.
\end{equation*}
In the specific climate context, some authors as \citet{LiLund2012} and \citet{Lu2010} use a MDL based-criterion (\citet{Rissanen78}). \citet{ardia2019frequentist} show that the MDL criterion can be seen as a Bayesian criterion with appropriate prior distributions for change-point models. As a consequence, the obtained based-MDL penalties (see \citet{LiLund2012,Lu2010}) looks like the mBIC (their both penalties integrate a term depending on the segment lengths of the segmentation).
\end{description}

\subsubsection{Different choices for our procedure} \label{sec:implementation}

The proposed inference procedure is summarized in Figure \ref{fig:algorithm} given in the Supplemental Material. The method is implemented in a R package named \texttt{GNSSseg} which is available on the CRAN. \\

In practice, Step 1 of the inference (Section \ref{sec:method:subsec:inference:subsub:inference_K}) is performed for $K=1,\ldots, K_{\text {max}}$ where $K_{\text{max}}$ should be 2 or 3 times larger than the expected number of change-points. For both the simulations and the applications, we used $K_{\text {max}}=30$.

The iterative procedure needs a proper initialization procedure and a stopping rule. For the initialization, the function $f$ is estimated first, using a unweighted least-squares criterion. For the stopping rule the change of $f_t$ and $\mu_k$ between two successive iterations is checked against a fixed threshold. The convergence of the iterative procedure is accelerated using the stopping test proposed by \cite{varadhan2008simple}. 

The final parameterization was derived after testing several different options which are described in the Supplemental Material.

\section{Simulation Study} \label{sec:sim}

\subsection{Simulation design and quality criteria.}\label{sec:sim:subsec:qc}

\subsubsection*{Simulation design.} 
\begin{figure}[ht]
\centering
\includegraphics[height=7 cm]{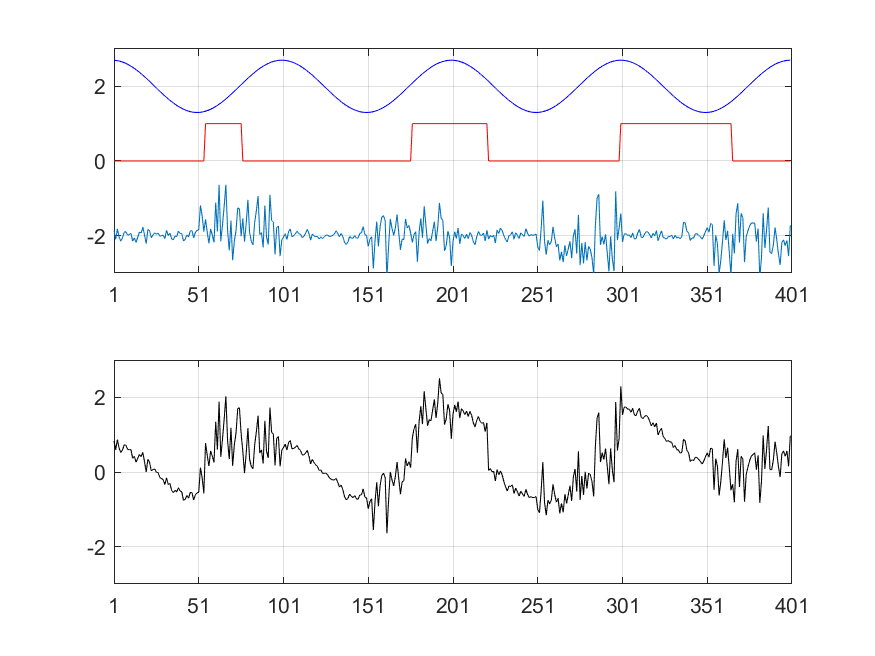}
\caption{Example of a simulated time series (black solid line in lower panel) of length $n=400$ with $K=7$ segments (red solid line), function $f_t=0.7 \cos(2 \pi t/L)$ (blue solid line), noise (cyan solid line) with standard deviation $\sigma_1^\star = 0.1$ and $\sigma_2^\star = 0.5$ (changing every $L/2=50$ points, starting with $\sigma_1^\star$).}\label{fig:simulation}
\end{figure}

The simulated time series are characterized by a length of $n=400$ with $4$ "years" of $2$ "months" of $50$ "days" each and with a monthly variance. A total of $6$ change-points were introduced at positions $t=55, 77, 177, 222, 300, 366$ and values for the signal mean were alternating between $0$ and $1$. The periodic function was modelled by $f_t=0.7 \cos(2 \pi t/L)$ where $L=100$ is the length of one year. Since we consider here only two months, the variance is alternating between two values, $\sigma_1^2$ and $\sigma_2^2$. Several batches of $100$ time series were generated with different values for $\sigma_1$ = $0.1$, $0.5$, or $0.9$ and $\sigma_2$ = $0.1$ to $1.5$ by step of $0.2$. Figure \ref{fig:simulation} shows an example.

\subsubsection*{Quality criteria.} The accuracy of the results is quantified by the differences between the estimates (denoted with a hat ${\widehat{x}}$) and the true values (denoted as $x^{\star{}}$).\\

For the function $f$, the root mean square error (RMSE) of the estimated function is computed: $\mbox{RMSE}(f) =\left[\frac{1}{n}\sum_{t=1}^{n} \{\hat{f}_t-f^{*}_t \}^2\right]^{1/2}$.

For the segmentation parameters, the following criteria are considered:
\begin{itemize}
    \item the difference between the estimated number of segments and the true one $\hat{K}-K^{*}$;
    \item the RMSE of the estimated mean parameter $\hat{\mubf}$: 
    $\mbox{RMSE}(\mubf)=\left[\frac{1}{n}\sum_{t=1}^{n}\left\{\hat{\mu}_t-\mu^{*}_t\right\}^2\right]^{1/2}$; 
    \item the distance between the estimated positions of the change-points $\widehat{\tbf}$ and the true ones $\tbf^{\star}$; this distance is measured with the help of the two components of the Hausdorff distance, $d_1$ and $d_2$, defined as: 
$$
d_1(a,b)=\max_b \min_a |a-b| \ \ \text{and} \ \ d_2(a,b)=d_1(b,a).
$$
%$d_1(\tbf^{\star}, \widehat{\tbf})$ quantifies the largest distance between an estimated change-point and the true ones. However, it does not say if some change-points are missing. This complementary information is given by $d_2 (\tbf^{\star}, \widehat{\tbf})$ which quantifies how close the true change-points are to the detected ones. 
A perfect segmentation results in both null $d_1(\tbf^{\star}, \widehat{\tbf})$ and $d_2(\tbf^{\star}, \widehat{\tbf})$. A small $d_1$ means that the detected change-points are well positioned and a small $d_2$ that a large part of the true change-points are correctly detected. A common situation found in practice is the one where the number of change-points is under-estimated, with a small $d_1$ and a large $d_2$. In that case, some change-points are undetected but the detected ones are correctly located. This situation is satisfying here since in our application it is acceptable to miss a few change-points (usually of small amplitude) rather than over-segmenting the data with badly-positioned change-points.

    \item the histogram of the change-point locations that provides a measure of the probability of the position of the change-points. 
\end{itemize}
%\end{description}

\subsection{Results.}\label{sec:sim:subsec:res}

Only the results for $\sigma_1^\star$ = $0.5$ are illustrated hereafter. The results for the others values of $\sigma_1^\star$ are briefly discussed at the end of the section.  

\subsubsection*{Accuracy of the variance estimates.} 
\begin{figure}[ht]
\centering
\includegraphics[height=5 cm, width=8cm]{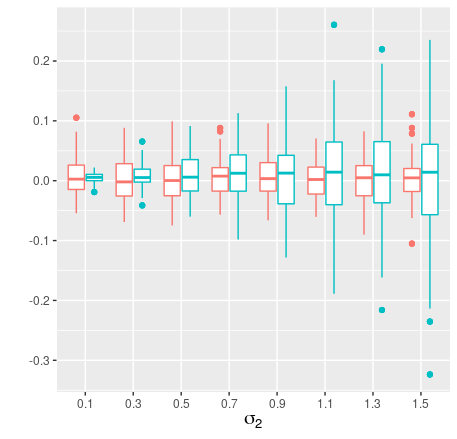}\\
\caption{Boxplots of standard deviation estimation errors: $\hat{\sigma}_{1}-\sigma_{1}^\star$ in red and $\hat{\sigma}_{2}-\sigma_{2}^\star$ in blue, with $\sigma_1^\star$=0.5 and $\sigma_2^\star=0.1,\dots,1.5$. Each case includes 100 simulations. }\label{fig:estimated_variance}
\end{figure}

Figure \ref{fig:estimated_variance} presents the estimation errors of $\hat{\sigma}_1$ and $\hat{\sigma}_2$ for different values of $\sigma_2^\star$. It is seen that the variance estimator works well and the estimated standard deviations are retrieved with the same accuracy as in \citet{Bock2018} despite the presence of the periodic bias. The dispersion increases when $\sigma_2^\star$ is increasing as one can expect.

\subsubsection*{Accuracy of segmentation parameter estimates.} 
\begin{figure}[h!]
\centering
\includegraphics[width=7cm]{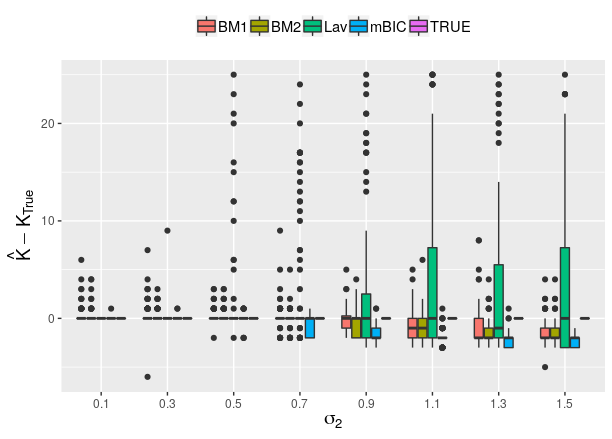}
\includegraphics[width=7cm]{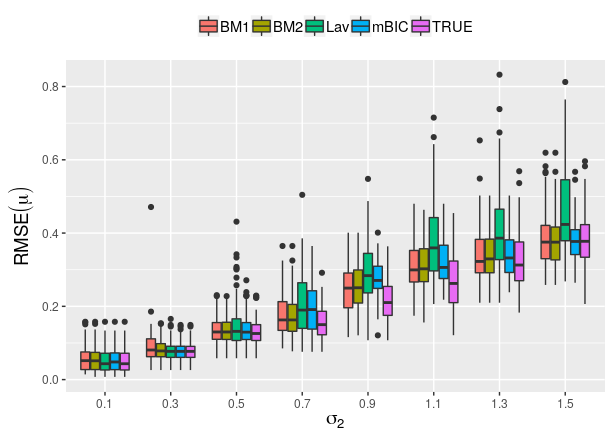}
\includegraphics[width=7cm]{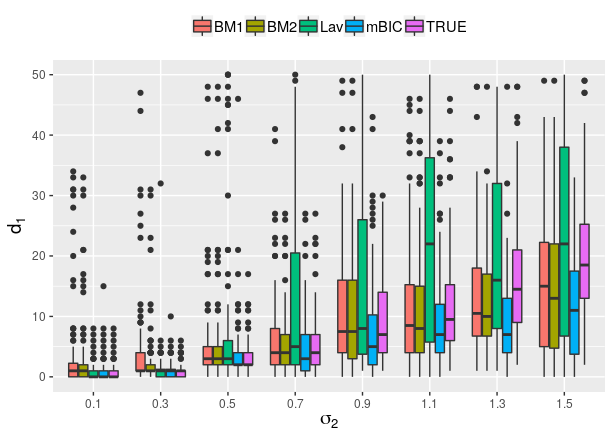}
\includegraphics[width=7cm]{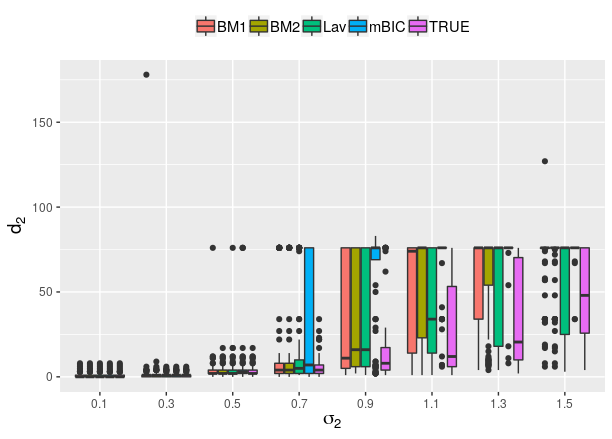}
\caption{Results with the four selection criteria (BM1, BM2, Lav, and mBIC) and with the true number of segments (True), for $\sigma_1^\star=0.5$ and different values of $\sigma_2^\star$. \emph{(a)} $\hat{K}-K^\star$; \emph{(b)} $\mbox{RMSE}(\mubf)$; \emph{(c)} first Hausdorff distance $d_1$ and \emph{(d)} second Hausdorff distance $d_2$.}
\label{fig:quality_criteria}
\end{figure}

Figure \ref{fig:quality_criteria} shows the results for the four model selection criteria and the special case where the number of segments $K$ is fixed to the true value ($K=7$). For small values of $\sigma_2^\star$, the detection problem is easy and all the model selection criteria retrieve the correct number of segments (Figure \ref{fig:quality_criteria}(a)). However for large values of $\sigma_2^\star$, the detection  becomes difficult, and the errors increase. The different criteria behave slightly differently. Lav tends to give the true number of segments in median, but with a large dispersion, while BM1, BM2, and mBIC tend to underestimate the number of segments (more for mBIC). However, finding the correct number of segments does not mean that the change-points are properly positioned. Indeed, for Lav and the case when $K=7$, the median $d_1$ is still quite large (Figure \ref{fig:quality_criteria}(c)). On the other hand, the median $d_2$ is smaller for the case when $K=7$ compared to the tested criteria (Figure \ref{fig:quality_criteria}(d)). Finally, RMSE($\mubf$) is very similar for all the criteria (Figure \ref{fig:quality_criteria}(b)), though Lav shows a larger median and dispersion when $\sigma_2^\star$ is large. When $\sigma_2^\star$ takes intermediate values the case when $K=7$ yields slightly improved results.

\subsubsection*{Probability of detection.} 
\begin{figure}
\centering
\includegraphics[scale=0.77]{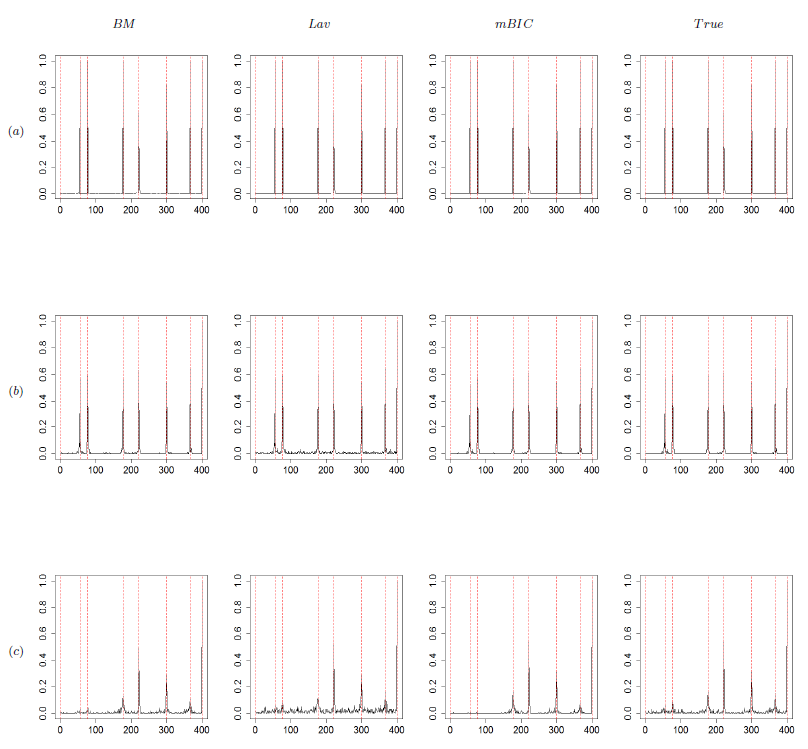}
\caption{Histogram of change-point detections with, from left to right, the BM1, Lav, and mBIC selection criteria, and the case when the true number of segments is used (TRUE), for $\sigma_1^\star=0.5$ and three different values for $\sigma_2^\star$: (a) $\sigma_2^\star=0.1$, (b) $\sigma_2^\star=0.5$ and (c) $\sigma_2^\star=1.5$. The red dotted lines indicate the positions of the true change-points. The results for BM2 are very similar to BM1 and are not shown.}
\label{fig:frequenze}
\end{figure}

Figure \ref{fig:frequenze} shows the percentage of the change-point detections for three values of $\sigma_2^\star=0.1, 0.5$ and $1.5$, and $\sigma_1^\star=0.5$. In general, the change-points located in the "months" with smaller variance are more often recovered with all three criteria, and also when the true $K$ is used. Hence, in the case (a) when $\sigma_1^\star=0.5$ and $\sigma_2^\star=0.1$, the probability of detection is slightly smaller for the position $222$, which is contained in a segment with $\sigma_1^\star = 0.5$, and for the position $300$ where both the mean and the variance change. In the case (b) when $\sigma_1^\star=\sigma_2^\star=0.5$, the probability of detection is more or less the same for all the change-points and all the criteria. When $\sigma_2^\star = 1.5$, the problem is more complicated. Again the change-points located in the "months" with smaller noise are better detected (positions $222$ and $300$) but for the other four change-points the results are contrasted although they are all located in months with $\sigma_2^\star=1.5$. The change-points at $55$ and $77$ are almost never detected. For mBIC this is consistent with the fact that the median $\hat{K}$=5, i.e. two change-points are missing, on average (Figure \ref{fig:quality_criteria}(a)), but the other four change-points are not so badly located ($d_1$ is not that large, Figure \ref{fig:quality_criteria}(c), but $d_2$ is very large, Figure \ref{fig:quality_criteria}(d)). The situation is a bit similar for BM1. On the other hand, for Lav and the true $K$, the number of detections is correct (on average for Lav) but due to the large noise they are sometimes very badly positioned (large $d_1$ and $d_2$).

\subsubsection*{Accuracy of the function estimate.} 
\begin{figure}[ht]
\centering
\includegraphics[width=8cm]{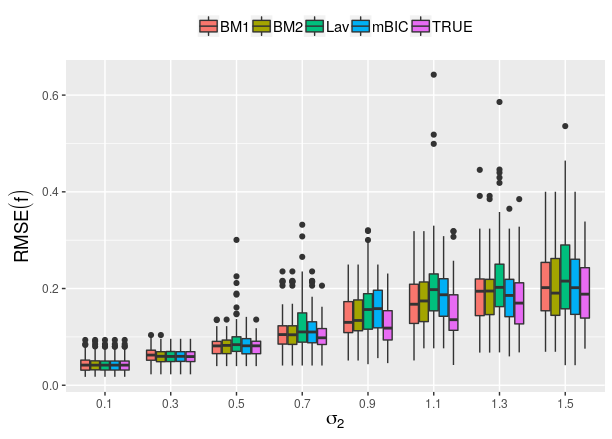}
\caption{RMSE of the estimated function $f$ for $\sigma_1^\star=0.5$ and different values for $\sigma_2^\star$.}
\label{fig:RMSE_f}
\end{figure}

Figure \ref{fig:RMSE_f} shows RMSE($f$) as a function of $\sigma^\star_2$. As expected, the errors increase when $\sigma_2^\star$ increases. The results do not much depend on the selection criterion, but the results are slightly better when the true number of segments is known and when $\sigma_2^\star$ takes intermediate values. The results for Lav show a slightly larger median and larger dispersion. \\

The results for other values of $\sigma_1^\star$ (not shown) are very similar for BM1, BM2, mBIC, and the case when the true $K$ is used. The results are slightly improved for $\sigma_1^\star=0.1$ and slightly degraded for $\sigma_1^\star=0.9$, as expected. The results for Lav are more chaotic, with either large under-estimation of $K$ for the smaller $\sigma_1^\star$ and over-estimation of $K$ for the larger $\sigma_1^\star$, with large subsequent degradation of the other quality criteria. In general, under-estimating $K$ leads to an increase of RMSE$(\mu)$, while over-estimating $K$ leads to an increase of $d_1$. \\

The main conclusions from the simulation study are the following:
\begin{itemize}
 \item The proposed method works well but the results are sensitive to the choice of the function form due to its possible confusion with the change-points. Performing a selection of the statistically significant parameters of the function appears as a good way to reduce this problem and improves slightly the change-point detection with our simulated data (see Supplemental Material). 
 \item Concerning the model selection criteria, BM1, BM2, and mBIC, provide very similar results. They behave well and detect correctly the number and position of change-points when the noise is not too large. When the noise is heavy some change-points are missed but this is a counterpart of the limited number of false detections. The Lav criterion shows much larger dispersion in the number of change-points and, though the estimated number is close to the truth in median, some change-points are not properly located (larger $d_1$ and $d_2$) with an impact on the estimated $\mubf$ and $f$.
\end{itemize}

\section{Application to real GNSS data}\label{sec:real}

\subsection{Dataset, metadata, and validation procedure}
The method is applied to the daily IWV differences from 120 global GNSS stations and ERA-Interim reanalysis for the period from 1 January 1995 to 31 December 2010. The dataset used here is the same as in \cite{Bock2019}. The metadata for the GNSS stations are available from the IGS site-logs (ftp://igs.org/pub/station/log/). They contain for each station the dates of changes of receiver (R), antenna (A), and radome (D). We also included the dates of processing changes (P) which occurred at a few stations in 2008 and 2009 (this issue is discussed in \citet{parracho2018global}). Experience shows that equipment changes do not produce systematically a break in the GNSS IWV time series. The most important changes are those affecting the antenna and its electromagnetic environment, the satellite visibility, and the number of observations \citep{Vey2009}. For instance, \citet{ning2016} considered only antenna and radome changes, as well addition/removal of microwave absorbing material which was known by the authors for one specific station. However, there is some evidence that changes in the receiver settings also induce inhomogeneities, e.g. when the elevation cutoff angle is changed. Changes in the environment due e.g. to cutting of vegetation and construction of buildings nearby the antenna as well as seasonal changes in multipath due to growing/declining vegetation may also impact the measurements and produce either abrupt or gradual changes. As a consequence, though metadata represent a valuable source of validation, a full matching between detected change-points and metadata is not to be expected. 

Because of noise in the signal, the detected changes may also not coincide perfectly with the known changes and we must allow some flexibility in the validation procedure. A window of 30 days before or after a documented change was used for the automatic validation of the detected change-points. A visual inspection was also performed to check if the invalidated change-points make sense. In some cases double detections just a few days apart are found on noise spikes, often with two large offsets of opposite signs. Such noise detections are classified as outliers.

\subsection{General Results}\label{sec:real:subsec:general_results}

\begin{figure}[ht]
\centering
{\includegraphics[width=16 cm]{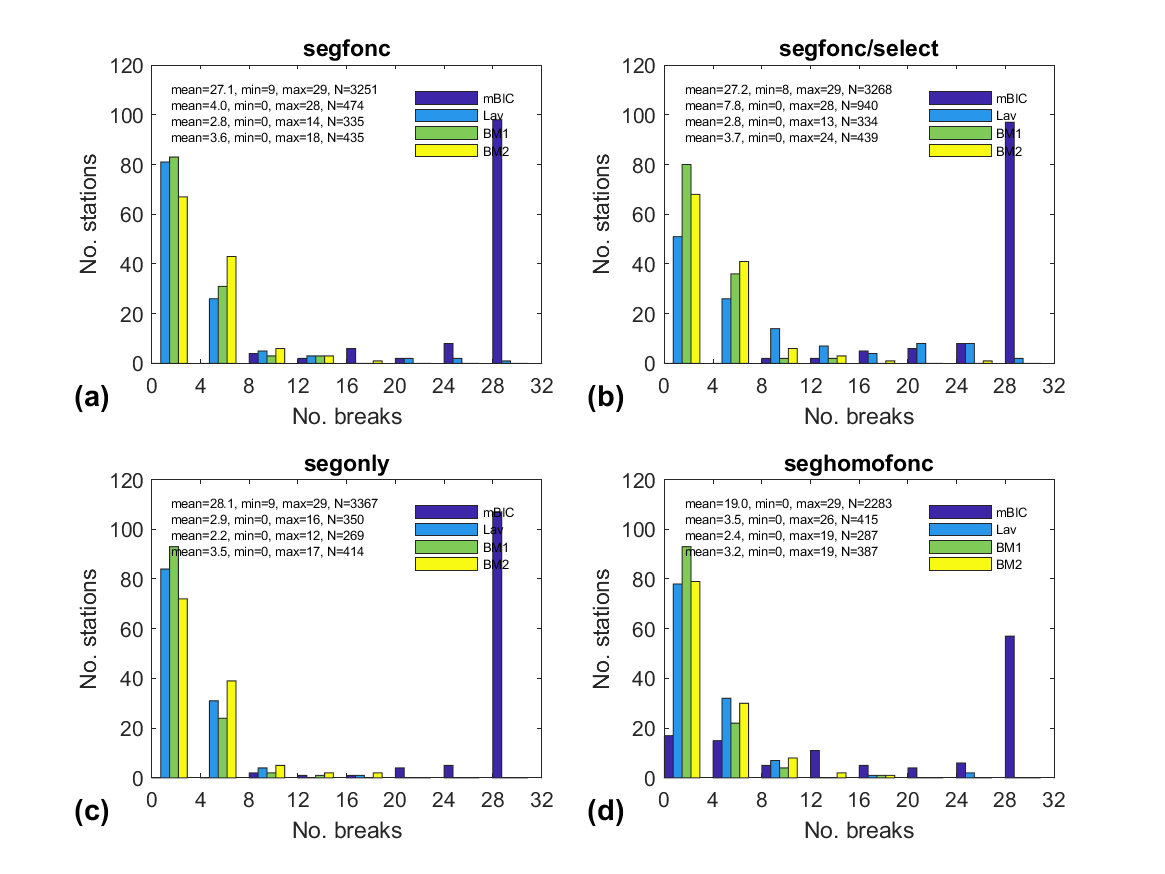}}
\caption{Histograms of the number of change-points detected for four variants of the model selecting criteria (mBIC, Lav, BM1, and BM2). The numbers given in the plots are the mean, min, and max number of change-points detected per station, N is the total number of change-points per method.\label{fig:stat_real}}
\end{figure}

In this section, we present results for the final method described in the preceding sections as well as for three alternative methods. The final method is referred to as variant (a). Variant (b) is a similar method where only the statistically significant terms of the Fourier series are selected. It is intended to check if reducing the number of degrees of freedom in the function leads to better results as was found with the simulations. Variant (c) is the earlier method proposed by \citet{Bock2018} in which only the segmentation is performed (i.e. the functional part is removed). Variant (d) is another form of a simplified method where the functional is modelled but a homogeneous variance is considered instead of a monthly variance. Statistics on the number of detected change-points are included in Figure \ref{fig:stat_real}. More statistics including the number of validations and outliers are given in Table \ref{table:criteres}.

\begin{table}[t]
\caption{Comparison of segmentation results for the four variants and the four model selection criteria. From left to right: Number of stations with change-points, min/mean/max number of detected change-points per station, total number of change-points, total number of outliers, total number of validations, percentage of validations including outliers, percentage of validations without outliers.}\label{table:criteres}
\begin{tabular}{llllllllll}
              &     &   &      &    &      &      &     &        &        \\ \hline
              & Nsta & min  & mean  & max & detections & outliers & validations \\ \hline
\multicolumn{10}{l}{\textbf{Variant (a)} (segfonc)}                                   \\ \hline
\textit{mBIC} & 120 & 9 & 27.1 & 29 & 3251 & 2096 & 267 & 8.2\%  & 20.9\% \\
\textit{Lav}  & 114 & 0 & 4.0  & 28 & 474  & 129  & 75  & 15.8\% & 21.3\% \\
\textit{BM1}  & 98  & 0 & 2.8  & 14 & 335  & 36   & 70  & 20.9\% & 23.3\% \\
\textit{BM2}  & 107 & 0 & 3.6  & 18 & 435  & 64   & 77  & 17.7\% & 20.6\% \\ \hline
\multicolumn{10}{l}{\textbf{Variant (b)} (segfonc/select)}                                   \\ \hline
\textit{mBIC} & 120 & 8 & 27.2 & 29 & 3268 & 2090 & 270 & 8.3\%  & 20.7\% \\
\textit{Lav}  & 115 & 0 & 7.8  & 28 & 940  & 411  & 116 & 12.3\% & 20.8\% \\
\textit{BM1}  & 100 & 0 & 2.8  & 13 & 334  & 46   & 68  & 20.4\% & 23.4\% \\
\textit{BM2}  & 107 & 0 & 3.7  & 24 & 439  & 76   & 81  & 18.5\% & 22.1\% \\ \hline
\multicolumn{10}{l}{\textbf{Variant (c)} (segonly)}                                   \\ \hline
\textit{mBIC} & 120 & 9 & 28.1 & 29 & 3367 & 1255 & 361 & 10.7\% & 16.4\% \\
\textit{Lav}  & 113 & 0 & 2.9  & 16 & 350  & 28   & 64  & 18.3\% & 19.6\% \\
\textit{BM1}  & 90  & 0 & 2.2  & 12 & 269  & 8    & 53  & 19.7\% & 20.2\% \\
\textit{BM2}  & 102 & 0 & 3.5  & 17 & 414  & 24   & 68  & 16.4\% & 17.4\% \\ \hline
\multicolumn{10}{l}{\textbf{Variant (d)} (seghomofonc)}                                   \\ \hline
\textit{mBIC} & 116 & 0 & 19.0 & 29 & 2283 & 1637 & 178 & 7.8\%  & 24.1\% \\
\textit{Lav}  & 114 & 0 & 3.5  & 26 & 415  & 148  & 56  & 13.5\% & 20.4\% \\
\textit{BM1}  & 92  & 0 & 2.4  & 19 & 287  & 40   & 61  & 21.3\% & 24.1\% \\
\textit{BM2}  & 101 & 0 & 3.2  & 19 & 387  & 82   & 68  & 17.6\% & 21.7\% \\ \hline
              &     &   &      &    &      &      &     &        &       
\end{tabular}
\end{table}

Figure \ref{fig:stat_real}(a) shows that with variant (a), mBIC, Lav, BM1, and BM2 detect a total of $3251$, $474$, $335$, and $435$ change-points, respectively. The distribution of the number of change-points per station is very different depending on the selection criterion. Most notably, mBIC detects between 9 and 29 change-points per station, with a mean value of $27.1$, i.e. in most cases the maximum number of segments is selected (here $K_{\text {max}}=30$). This behaviour was not observed with the simulations. From Table \ref{table:criteres} we see that mBIC detects many outliers. Comparison of contrast values reveals that mBIC selects solutions with smaller SSR values than the other criteria, i.e. the model selected by mBIC generally explains better the observed signal. However, this is at the expense of strong over-segmentation, which is not wanted. mBIC is thus not well adapted to the nature of the data analyzed here. One of the reasons might be that the hypothesis of Gaussian errors is not valid (e.g. due to serial correlation in the data and noise spikes). The three other selection criteria provide much more consistent results, with mean number of change-points of $2.8$, $3.6$ and $4.0$ for BM1, BM2, and Lav, respectively. Among the three criteria, we see from Table \ref{table:criteres} that BM1 has the smallest number of outliers (36) and the highest rate of validations ($20.9 \%$). These two features, and also the fact that BM1 has a reasonable number of change-points (the mean is 2.8 per station), make this selection criterion the preferred one.

Compared to variant (a), variant (b) shows marginal impact on the number of detections and the number of validations for three criteria (mBIC, BM1, and BM2). Only for Lav do the mean and total number of detections increase (by nearly a factor of 2). This behavior is not explained but it reveals some instability in the model selection with this criterion. Instability could also be guessed from the maximal number of detections of 28 already seen in variant (a). It means that in some cases, Lav selects a number of segments very close to the maximum ($K_{\text {max}}=30$). BM1 and BM2 have also more outliers with this variant, though the total number of detections is almost unchanged. So, contrary to the simulation results, with the real data there is no benefit of applying a selection of significant terms of the functional model.

In variant (c), the result for mBIC is slightly worse (more detections) but with fewer outliers. For the three other criteria the number of detections decreases significantly. The latter behaviour was actually not expected. Our interpretation is that when the periodic bias is not modelled, the segmentation algorithm has two options: either (i) put additional change-points to better fit the periodic variations in the signal, but this would lead to many more detections (4 per year, i.e. a total of 64 per station for a 16-year time series), or (ii) select only those change-points with a large amplitude that are not confounded with the periodic bias. The observed result (Figure \ref{fig:stat_real}(c) and Table \ref{table:criteres}) suggest that BM1, BM2, and Lav select the second, more conservative, option. Our final method is actually capable of detecting smaller offsets, which makes it more efficient for the homogenization purpose. Note that with variant (c), the situation described by option (i) occurs nevertheless in some cases, as will be illustrated in the next sub-section, and though the number of outliers and validations both decrease for BM1, BM2, and Lav, the percentage of validations remains nearly the same (Table \ref{table:criteres}). So, variant (a) clearly works better than variant (c) in the sense it detects more change-points; it has nevertheless the drawback of detecting more outliers. This point is further discussed in the last section.

In variant (d) the variance is assumed to be constant. This has two consequences: (i) the function is fitted with uniform weights which in general leads to an estimated function $\hat{f}$ and an estimated mean $\hat{\mu}$ of different shapes, (ii) the estimated variance is larger than the mean variance of the variant (a) (the average mean standard deviations amount to $1.19$ vs. $0.84 kgm^{-2}$, respectively) and fewer change-points are detected. Table \ref{table:criteres} confirms that with this method fewer change-points are detected than with variant (a), however the number of outliers is increased (except for mBIC which is again a special case). The number of validations is also decreased, but the percentage of validations is almost unchanged. \\

\begin{figure}[ht]
\centering
{\includegraphics[width=16 cm]{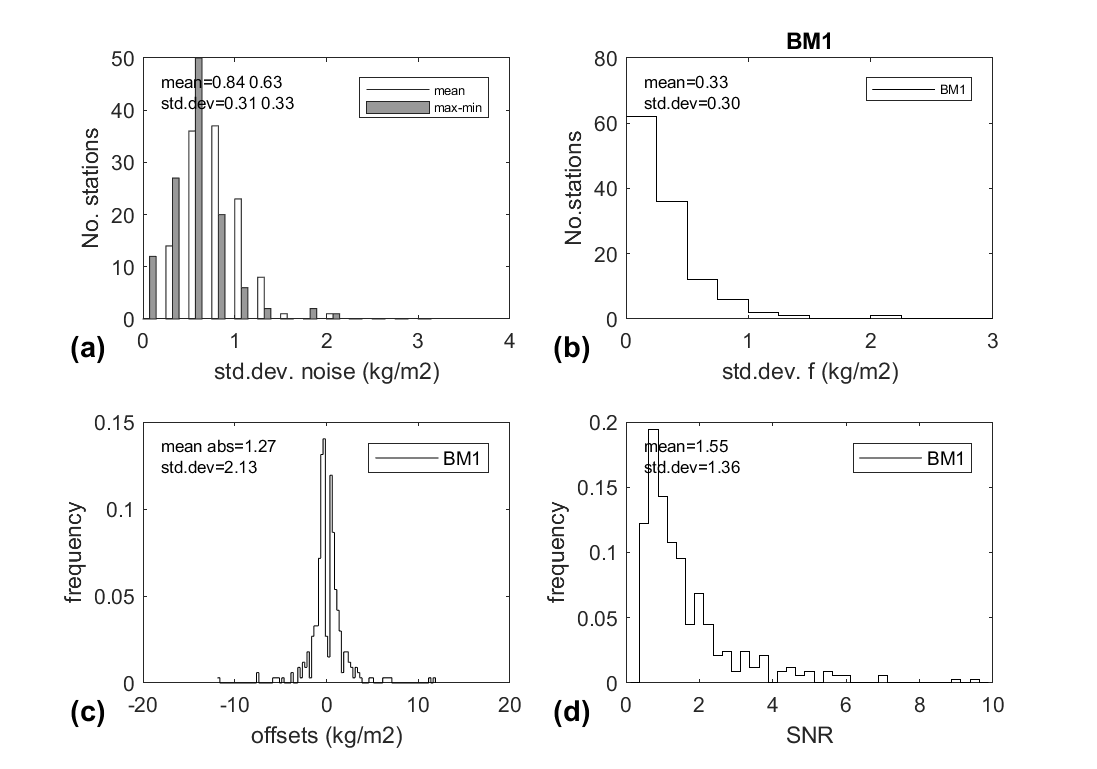}}
\caption{Histograms of segmentation results for the final method with selection criterion BM1: (a) Number of stations with respect to the estimated standard deviation of the noise (mean and max-min of the 12 monthly values); (b) Number of stations with respect to the standard deviation of the estimated function; (c) Distribution of offsets of detected change-points; (d) Distribution of SNR of detected change-points.\label{fig:offsets}.}
\end{figure}

The comparison of the four variants shows thus that the final method, including a heterogeneous variance and a full functional model for the periodic bias, has the best properties: reasonable number of detections and outliers, and high rate of validations. Among the four model selection criteria, BM1 and BM2 behave better than Lav and mBIC, with a small advantage for BM1. Figure \ref{fig:offsets}(a) shows that the yearly-mean standard deviation of the noise ranges between 0 and 2 $kg m^{-2}$, with a mean value over the 120 stations of $0.84 kg m^{-2}$. The seasonal excursion is of $0.63 kg m^{-2}$ on average, which reflects the importance of modelling the heterogeneous variance. Figure \ref{fig:offsets}(b) presents a measure of the magnitude of the periodic bias for BM1. With an average value of 0.33 $kg m^{-2}$ it is clear that the periodic bias is not negligible and modelling it improves the segmentation results as shown by comparing the results of variant (d) and (a). Figure \ref{fig:offsets}(c) shows that the distribution of offsets (changes in mean) is nearly symmetrical. The mean absolute value of $1.27 kg m^{-2}$ is relatively large. The dip centred on zero reflects the fact that the smaller offsets are more difficult to detect because of their small signal-to-noise (SNR) ratio. The most frequently detected offsets are found around +/- $0.5 kg m^{-2}$. The larger offsets (up to +/- $10 kg m^{-2}$) are outliers. The distribution of SNR can be computed as the absolute value of offset divided by standard deviation of noise. It is peaking at 0.6 and the larger values (up to 10) correspond again to outliers (Figure \ref{fig:offsets}(d)). The mean SNR of 1.55 indicates that our method has a good efficiency of detection.

\subsection{Examples of special cases}

In addition to the global results, we exhibit the results for four stations showing for special cases of the variants. Only the criterion BM1 is considered here. With variant (c) there are actually 66 stations which have the same number of detections as variant (a). Though in general the change-points are located at the same position in the time series, this is not always the case. For 18 stations, variant (c) detects more change-points and for 36 stations it detects fewer. Station POL2 is an example of the former category and station STJO an example of the latter. DUBO is an example where the same number is detected but the change-points are not located at the same position. With variant (d), the number of stations with equal, more, and fewer numbers of detections is: 57, 24, and 39, respectively. Examples are: EBRE, MCM4, and POL2, respectively. \\

\begin{figure}[ht]
\centering
{\includegraphics[height=16 cm]{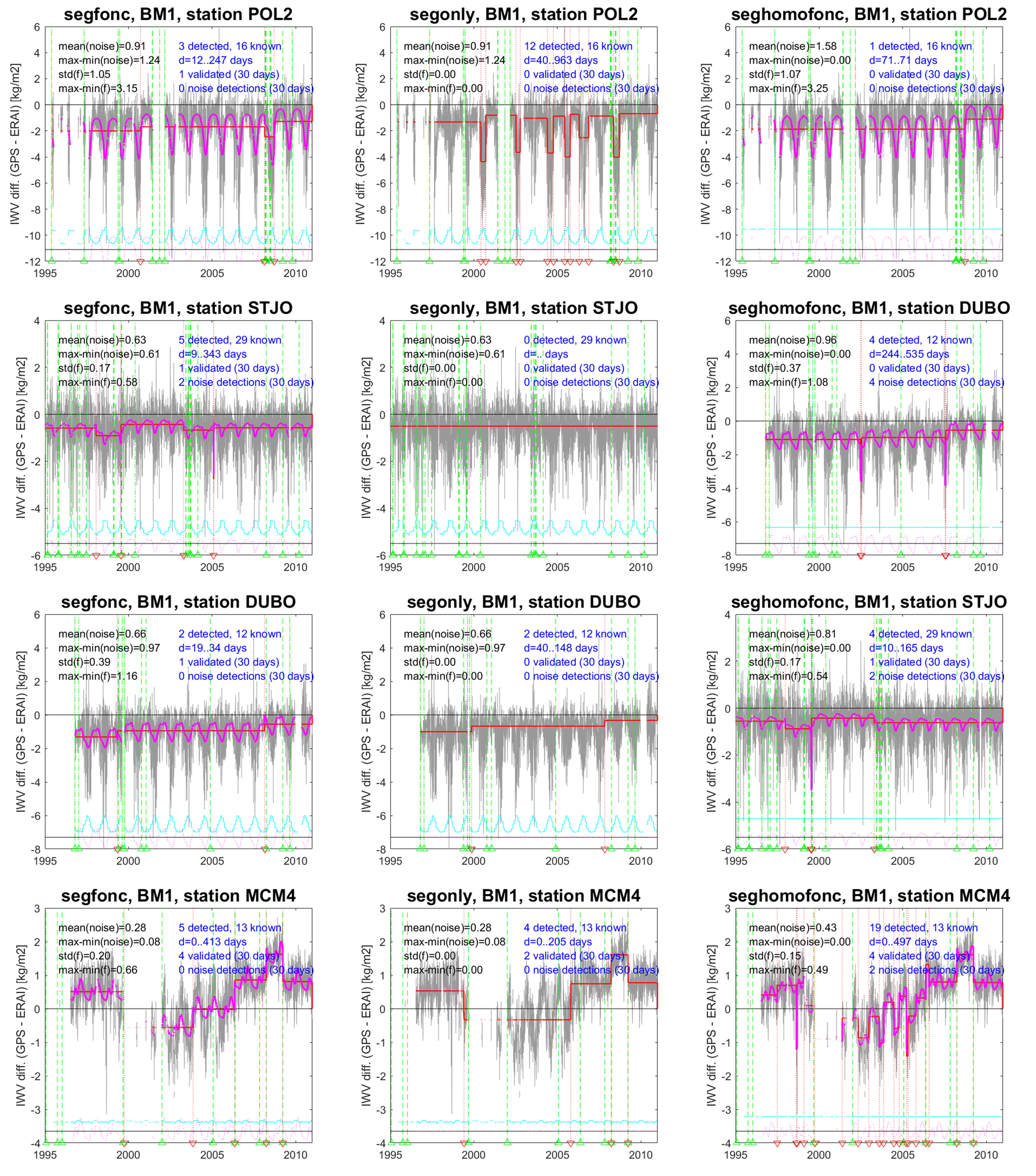}}
\caption{Examples of results obtained with variants (a), (c), and (d) from left to right, for four different stations: POL2, STJO, DUBO, and MCM4 (from top to bottom). The content of the plots is similar to Fig. \ref{fig:ccjm}(b). The text inserted at the top left of the plots reports the mean standard deviation of the noise, the variation (max-min) of the standard deviation of the noise, the standard deviation of the periodic bias function, and the variation (max-min) of the periodic bias function. The text in blue reports the total number of detections and of known changes, the minimum and maximum distance between detected change-points and the nearest known changes, the number of validated detections, and the number of noise detections. \label{fig:special_cases}}
\end{figure}

The results for a selection of four stations are given in Figure \ref{fig:special_cases}:
\begin{itemize}
\item In the case of POL2, variants (a), (c) and (d) detect 3, 12, and 1 change-point, respectively. The signal shows a strong periodic variation which well fitted by the models of variant (a) and (d) but is erroneously captured by the segmentation in variant (c). Variant (a) has one validated change-point (detected date: 2008-02-23, known change: 2008-03-06, type of change: P). Variant (c) has no validation, although it detects 12 change-points.  Variant (d) detects only one change-point, which is located 72 days from the nearest known change-point and is thus not validated, but it coincides with one of the three detections found by variant (a). The detection of this change-point is made difficult because it is located in a month with heavy noise. 
\item In the case of STJO, variants (a) and (d) detect 5 and 4 change-points, respectively, with one outlier each but not at the same position. Among the detected change-points, one is exactly the same (detected: 2003-04-18, known: 2003-06-08, type: R) but is not validated, and one is close (detected by variant (a): 1999-07-20, by variant (d): 1999-07-19, known: 1999-07-29, type: R) and is validated. Variant (c) gives no detection, the conservative option is selected by BM1 (option (ii) discussed above). 
\item In the case of DUBO, variants (a) and (c) detect two change-points at almost the same position but not exactly. Both are located close to known changes but only one is validated for variant (a) (detected: 1999-05-07, known: 1999-05-26, type: R). The second one is located 34 days from a known change for variant (a) and 148 days for variant (c). Though variant (c) works not bad, it is not as accurate as variant (a) because the periodic bias is neglected. Variant (d) has 4 detections which actually consist in 2 change-points, each being associated with an outlier. Although the periodic bias is modelled here, both change-points are quite badly located and thus not validated.
\item Finally for MCM4 the signal has very marked inhomogeneities in the form of several abrupt changes but also non-stationary oscillations. The abrupt changes are well captured by variant (a) who detects 5 change-points among which 4 are validated (types are in chronological order: R, R, P, P). The non-stationary oscillations are only partly modelled by the periodic function. This is a special case where even the model used in variant (a) is not well adapted to such oscillations. This result advocates for an improvement of the functional basis. In that case, variant (c) works quite well too and leads to almost the same detections as variant (a), but only the two P changes are validated. Variant (d) on the other hand over-estimates the number of change-points to better fit the non-stationary oscillations but with detections of outliers. The four same change-points are validated as with variant (a) but the fitted means are quite different.
\end{itemize}

Among the 70 validated change-points found by BM1 in the case of variant (a) there are 53 R, 16 A, 7 D, and 13 P types (note that these numbers don't sum up to 70 because in many cases the changes involve several types). We find here that receiver changes are the most frequent explanation for inhomogeneities. This is not surprising since they are the most frequent change-type occurring at GNSS stations. However, this is in contrast with \citet{ning2016}'s results who did not consider receiver changes at all. About 70\% of the receiver changes documented in the IGS sitelogs actually refer to firmware updates which don't have much impact on the observations as long as they don't involve a change in the minimum elevation cutoff angle. Hardware changes on the other hand are more prone to have an impact. We performed a quality control based on the observation files with TEQC software \citep{esteymertens1999} and found that in many cases hardware changes lead to changes in the multipath diagnostic parameters and in some occasions in the percentage of observations. Receiver changes that have an impact are e.g. found at station STJO on 1999-08-06 (from ROGUE\_SNR\_8000 to AOA\_SNR\_12\_ACT) and at station MCM4 on 2002-01-03 (from ROGUE\_SNR\_8000 to AOA\_SNR\_12\_ACT) and on 2006-05-19 (from AOA\_SNR\_12\_ACT to ASHTECH\_ZXII3). At MCM4, strong oscillations are found in the multipath diagnostics (mp1 and mp2) during the AOA\_SNR\_12\_ACT period, similar to those seen in the IWV differences (Figure \ref{fig:special_cases}). This reveals a malfunctioning of the GNSS equipment also associated with a jump in the mean signal at the beginning and at the end of that period.

\subsection{Comparison with \citet{ning2016}}

Similar to this study, \citet{ning2016} analyzed the homogeneity of GNSS-ERAI IWV differences for a global network of 101 GNSS sites with a least 15 years of observations. Their series were used with monthly sampling whereas here we used daily sampling. They used the PMTred test \citep{Wang2008} to detect abrupt changes in the mean IWV difference but this model does neither include a periodic bias not a monthly varying variance. They detected a total of 62 change-points affecting 47 stations among which 45 detections were attributed to the GNSS series, 16 to ERAI, and 1 was undetermined. Their attribution method was based on the comparison of the GNSS candidate series to two or three references series (ERAI, another nearby GNSS series, and/or a nearby VLBI series). Consistency between the two or three detected offsets was used to attribute the change-points to GNSS and disagreement to ERAI (by default). They also validated 13 detections with the GNSS metadata, but they included only antenna, radome, and known microwave absorbing material changes. Their validation window was +/- 6-month wide, i.e. much larger than our +/- 30-day window. We reviewed their validations for 42 of their sites for which we had metadata information from the IGS sitelogs including in our case receiver changes. Using the same 6-month window, we found that 10 out of their 12 undocumented GNSS detections can actually be explained with receiver changes and 2 with receiver+antenna changes (the latter were surprisingly missing in their analysis). Six of these changes agreed with the metadata within 2 months or less. We also found that 5 out of 15 of their  change-points attributed to ERAI coincide actually with 2 GNSS receiver changes and 3 antenna changes. Finally, inspection of the GNSS-ERAI IWV difference time series suggests that many of their undocumented detections may be due to outliers and gaps in the time series. This suggests that the implementation of the PMTred test is quite sensitive to fluctuations in the noise, a property similar to that of variant (d) discussed in the previous sub-section.

The comparison of our results for variant (a) with \citet{ning2016}'s results for 31 common stations which have change-points leads to the following conclusions: (i) our method detects nearly twice more change-points than PMTred (107 vs. 43), (ii) among 32 PMTred detections attributed to GNSS, about 1/3rd coincide with ours within +/- 2 months, 1/3rd within 2-6 months and 1/3rd within more than 6 months, (iii) among 11 PMTred detections attributed to ERAI, 4 change-points coincide with ours within +/- 1 month (the others being about 6 months or more apart) and none of them can actually be explained by GNSS changes (even involving receiver changes). Inspection of the IWV differences and the TEQC diagnostics confirms that the 4 change-points attributed to ERAI cannot be explained by changes in the GNSS time series, i.e. they may truly be due to ERAI; these are: GODE (1998-08-06), HOB2 (2006-06-10), and WUHN (1999-02-14 and 2006-09-27). The latter change-point was already mentioned by \citet{parracho2018global} as being due to a change in radiosonde data from the station at the city of Wuhan, China, being assimilated in ERAI.

\section{Discussion and conclusions} 
\label{sec:conclusion}
In this paper we presented a new segmentation method for the detection of abrupt changes in the mean of geophysical time series including a periodic bias and heterogeneous variance. The results on simulated data showed that the segmentation results (position and amplitude of change-points) are sensitive to the choice of the function basis used to model the periodic bias and to the initialisation of the iterative procedure in which the function and segmentation parameters are estimated. Several model selection criteria were tested. The criterion proposed by \citet{BM2001} and the modified BIC proposed by \citet{Zhang2007} appeared to have good properties. The criterion of \citet{Lavielle2005} appears rather unstable with large dispersion in the number of detected change-points.

When applied to real data (GNSS minus ERAI IWV series), the modified BIC's results were very disappointing (strong over-estimation of the number of change-points), certainly due to the fact that it is derived in the case of a normal distribution and a homoscedastic variance case. In fact, all the considered model selection criteria are based on these assumptions, but according to our experience mBIC is much more sensitive to deviations from the normal distribution.

We tested several variants of the method with the real data and found that accounting for a monthly variance and a period bias improved clearly the detection, although this method has some tendency to detect outliers due to noise spikes (about 20\% of the detections). A proper outlier detection method has to be developed, e.g. based on the SNR, in order to reject these detections.  

In addition, future improvements of the proposed method would be: (i) to consider other models for the function $f$ since it was found that in some cases like at station MCM4 a simple periodic function is not adequate, (ii) to take the serial correlation in the data into account. The first point can be handled by an estimation of the function $f$ using a non-parametric approach. The second point can be developed by following the approach of \citet{Chakar2017} who proposed to model the temporal correlation using an autoregressive process of order 1. These authors also proposed a two-stage whitening inference strategy that allows the use of the DP algorithm and find the exact maximum likelihood solution.

%%%%%%%%%%%%%%%%%%%%%%%%%%%%%%%%%%%%%%%%%%%%%%%%%%%%%%%%%%%%%%%%%%%%%
% DATA AVAILABILITY STATEMENT
%%%%%%%%%%%%%%%%%%%%%%%%%%%%%%%%%%%%%%%%%%%%%%%%%%%%%%%%%%%%%%%%%%%%%
\section*{\datastat}
The GNSS IWV data are available from \href{}{https://doi.org/10.14768/06337394-73a9-407c-9997-0e380dac5591}. (last access: April 2020; \citep{Bock2016}). 

ERA-Interim data are avaialable from \href{}{https://www.ecmwf.int/en/forecasts/datasets/archive-datasets/reanalysis-datasets/era-interim} (last access: April 2020; \citep{Dee2011}).

%%%%%%%%%%%%%%%%%%%%%%%%%%%%%%%%%%%%%%%%%%%%%%%%%%%%%%%%%%%%%%%%%%%%%
% ACKNOWLEDGMENTS
%%%%%%%%%%%%%%%%%%%%%%%%%%%%%%%%%%%%%%%%%%%%%%%%%%%%%%%%%%%%%%%%%%%%%
\section*{\ackname}
This work was developed in the framework of the VEGA project and supported by the CNRS program LEFE/INSU. This paper is IPGP contribution number 4136. The contribution of the third author has been conducted as part of the project Labex MME-DII (ANR11-LBX-0023- 01) and within the FP2M federation (CNRS FR 2036).

%%%%%%%%%%%%%%%%%%%%%%%%%%%%%%%%%%%%%%%%%%%%%%%%%%%%%%%%%%%%%%%%%%%%%
% REFERENCES
%%%%%%%%%%%%%%%%%%%%%%%%%%%%%%%%%%%%%%%%%%%%%%%%%%%%%%%%%%%%%%%%%%%%%

\newpage
\section*{Supplemental Material} \label{sec:supplement}

\subsection*{Summary of the proposed procedure}
\begin{figure}[ht]
%\centering
\caption{Schematic of the algorithm.}
\includegraphics[scale=0.6]{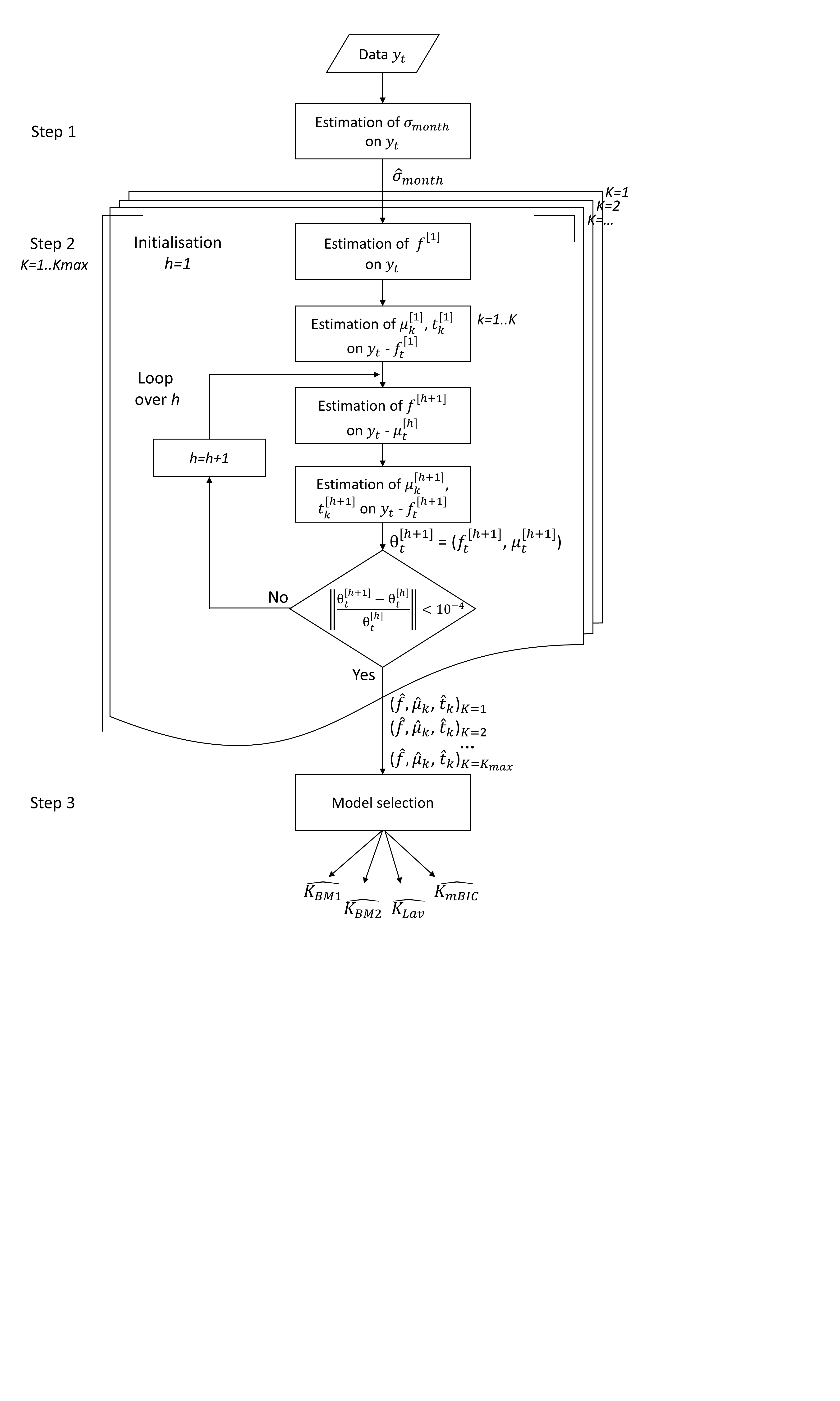}
\label{fig:algorithm}
\end{figure}

Figure \ref{fig:algorithm} summarizes the proposed procedure. 

\subsection*{Tested alternatives to the proposed procedure}

\begin{figure}[ht]
\centering
\includegraphics[height=5 cm, width=8cm]{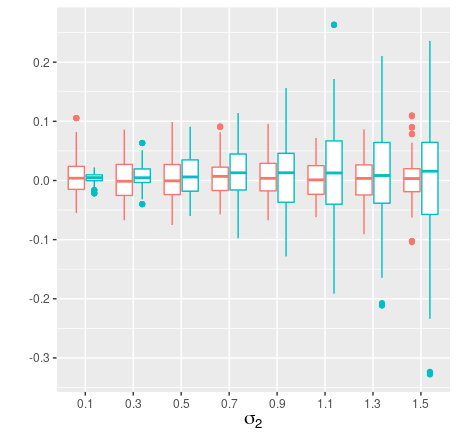}
\caption{Boxplots of standard deviation estimation errors for variant (1): $\hat{\sigma}_{1}-\sigma_{1}^\star$ in red and $\hat{\sigma}_{2}-\sigma_{2}^\star$ in blue, with $\sigma_1^\star$=0.5 and $\sigma_2^\star=0.1,\dots,1.5$. \label{fig:S1}} 
\end{figure}

\begin{figure}[ht]
\centering
\includegraphics[width=7cm]{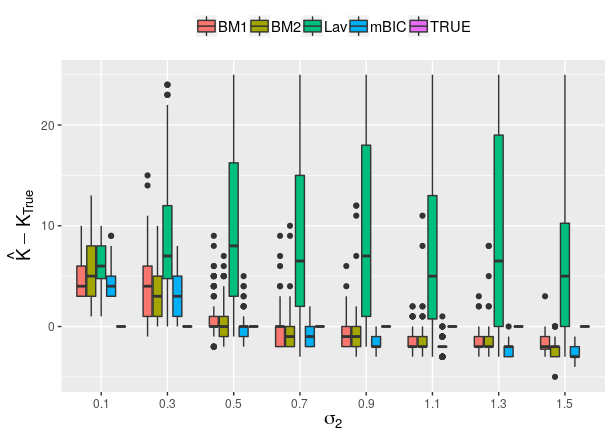}
\includegraphics[width=7cm]{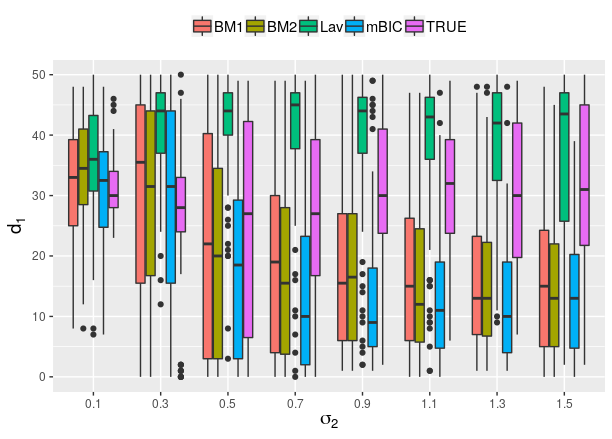}
\includegraphics[width=7cm]{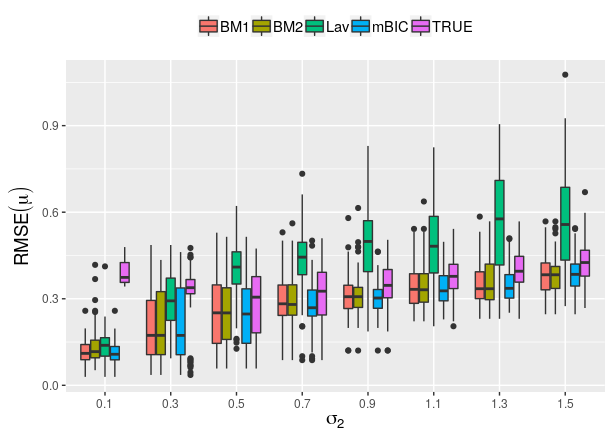}
\includegraphics[width=7cm]{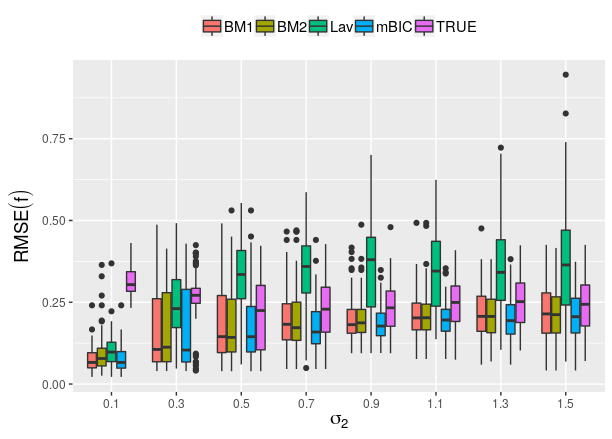}
\caption{Results for variant 2-(a). \emph{(a)} $\hat{K}-K^\star$; \emph{(b)} first Hausdorff distance $d_1$; \emph{(c)} $\mbox{RMSE}(\mubf)$; \emph{(d)} $\mbox{RMSE}(f)$.}
\label{fig:tot_out_init_seg}
\end{figure}

\begin{figure}[ht]
\centering
\includegraphics[width=7cm]{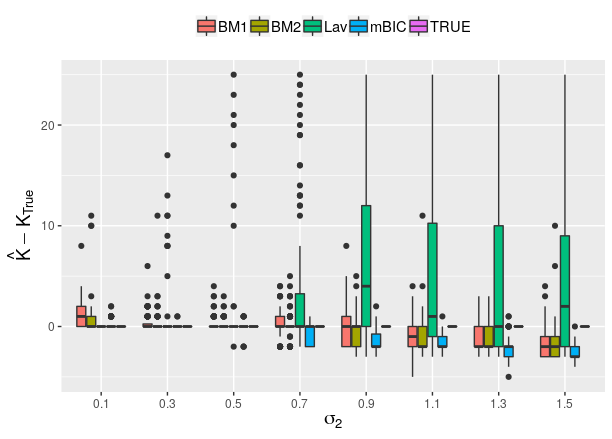}
\includegraphics[width=7cm]{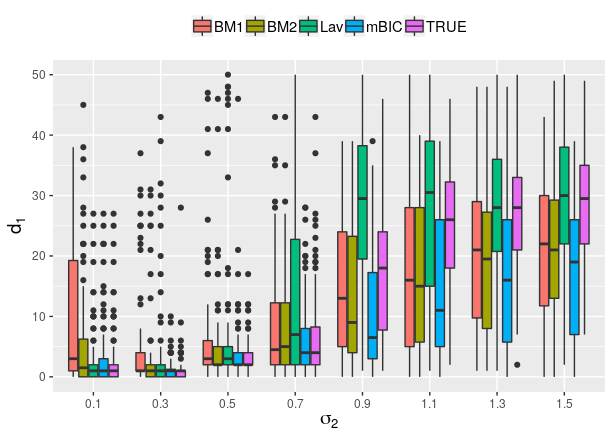}
\includegraphics[width=7cm]{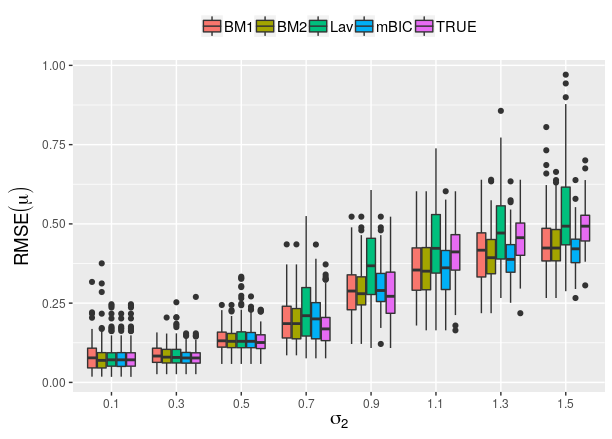}
\includegraphics[width=7cm]{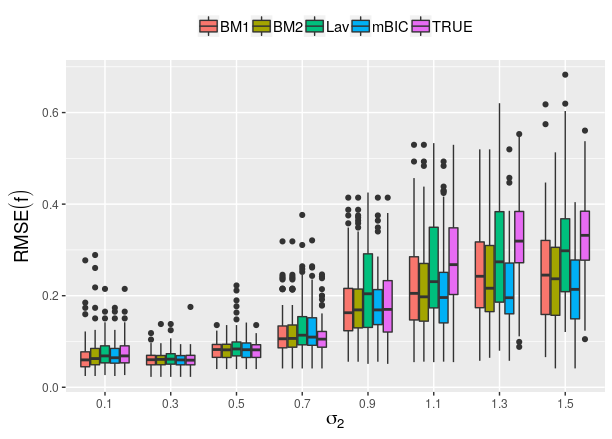}
\caption{Results for variant 2-(b). \emph{(a)} $\hat{K}-K^\star$; \emph{(b)} first Hausdorff distance $d_1$; \emph{(c)} $\mbox{RMSE}(\mubf)$; \emph{(d)} $\mbox{RMSE}(f)$.}
\label{fig:tot_out_init_weighted}
\end{figure}

\begin{figure}[ht]
\centering
\includegraphics[width=7cm]{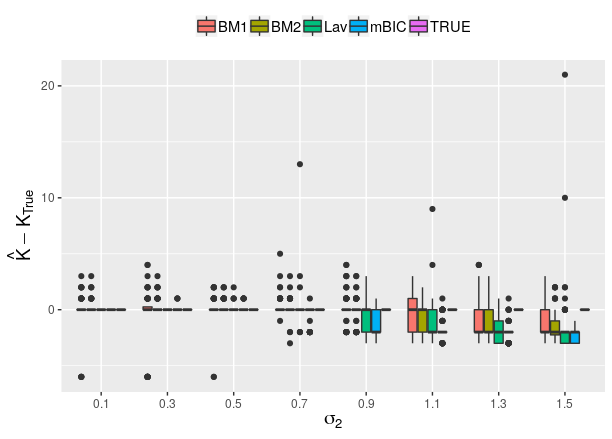}
\includegraphics[width=7cm]{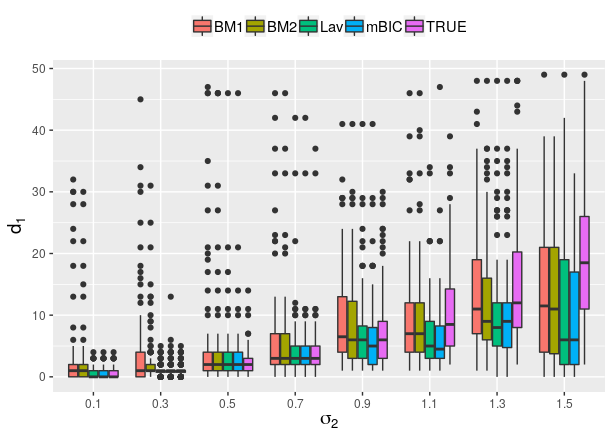}
\includegraphics[width=7cm]{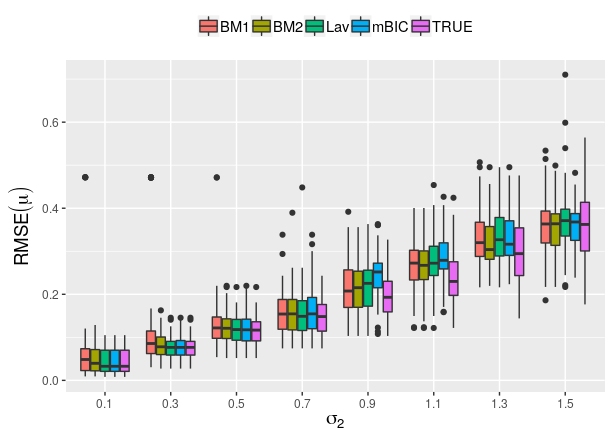}
\includegraphics[width=7cm]{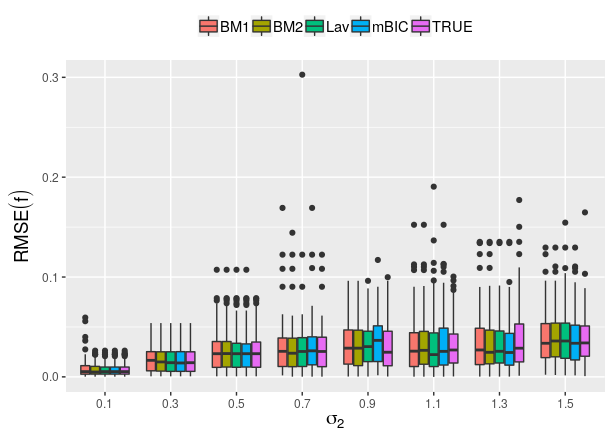}
\caption{Results for variant 3-(a). \emph{(a)} $\hat{K}-K^\star$; \emph{(b)} first Hausdorff distance $d_1$; \emph{(c)} $\mbox{RMSE}(\mubf)$; \emph{(d)} $\mbox{RMSE}(f)$.} 
\label{fig:cos_out_init}
\end{figure}

\begin{figure}[ht]
\centering
\includegraphics[width=7cm]{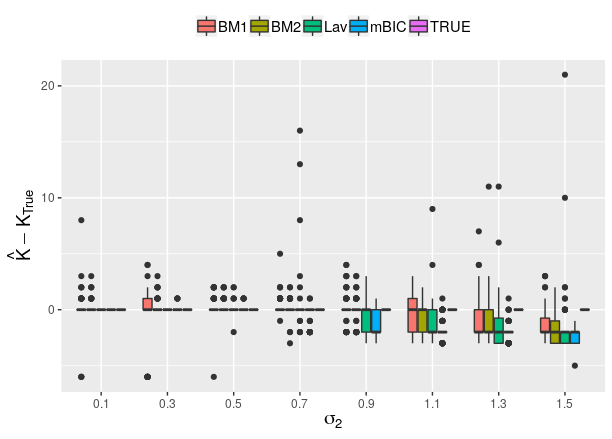}
\includegraphics[width=7cm]{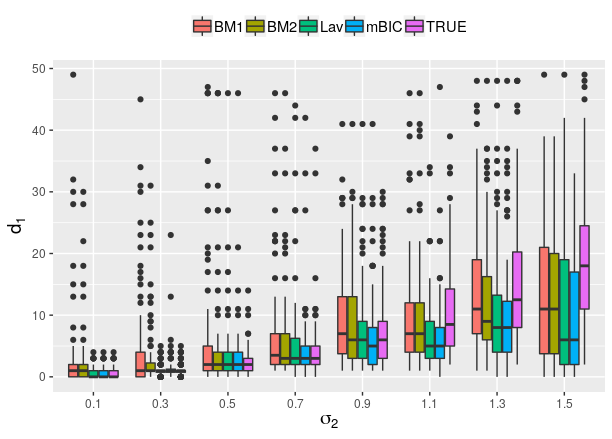}
\includegraphics[width=7cm]{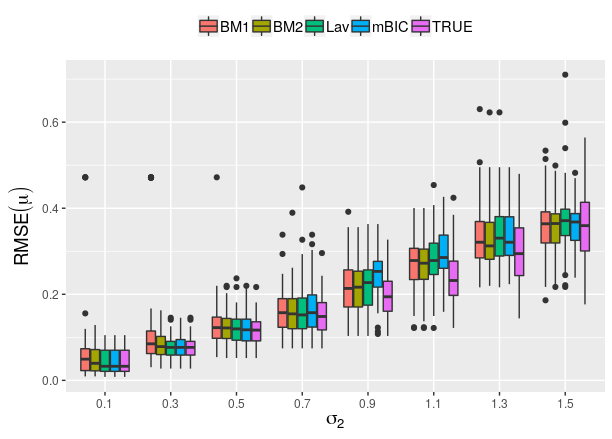}
\includegraphics[width=7cm]{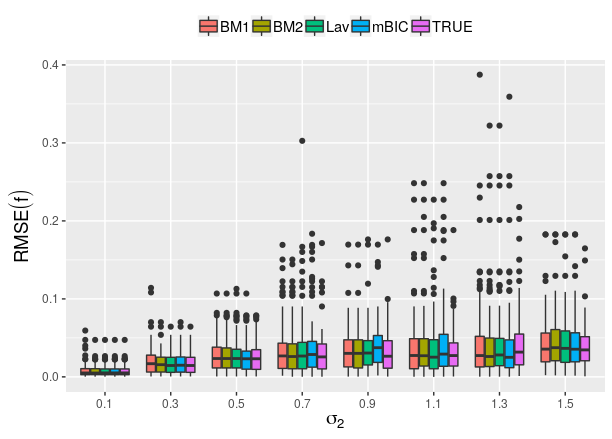}
\caption{Results for variant 3-(b). \emph{(a)} $\hat{K}-K^\star$; \emph{(b)} first Hausdorff distance $d_1$; \emph{(c)} $\mbox{RMSE}(\mubf)$; \emph{(d)} $\mbox{RMSE}(f)$.}  
\label{fig:selb_out_init}
\end{figure}

Recall that in our procedure (see Section \ref{sec:method:subsec:inference}), (1) the variances are estimated first; (2) the iterative procedure is initialized by the estimation of $f$ using an unweighted leas-square criterion; (3) the function is estimated with a Fourier decomposition of order $4$. We tested different variants for these three points:

\begin{description}
 \item[(1) Updating the variances] we tested a version of the procedure where ${\bf\sigma}$ was updated at each iteration of the iterative procedure. The estimated variances are plotted in Figure \ref{fig:S1}. This option provided slightly more accurate estimates for all the variance (see Figure \ref{fig:estimated_variance}) and the function parameters with very little impact on the segmentation parameters (not shown) compared to our procedure. However, the small changes in variance at each iteration severely slowed down the convergence of the algorithm.  \\
 
 \item[(2) Variants of the initialization] three variants are tested: (a) the segmentation is performed first; (b) $f$ is estimated first using a weighted regression (as in the iterative procedure); (c) $f$ is estimated first using a weighted regression but on the centered signal $y_t - \bar{y}$. 
 
 Figure \ref{fig:tot_out_init_seg} shows the results for option (a). Compared to the results of our procedure (see Figs. \ref{fig:quality_criteria} and  \ref{fig:RMSE_f}), the results are significantly degraded. Especially, the larger $d_1$ indicates that change-points are badly located. At the beginning, the unmodelled periodic variations present in the signal are captured by the segmentation. The iterative procedure does not change this effect leading naturally to an over-segmentation  in addition of the bad estimation of $f$. This is particularly marked for small values of the noise $\sigma_2$ and for the Lav criterion whatever $\sigma_2$. 
 
 Figure \ref{fig:tot_out_init_weighted} shows the results for option (b). The results are degraded as well but less than previously and mainly for larger $\sigma_2$. This can be explained by the fact that the unmodelled change-points belonging to small variance periods are absorbed by $f$ degrading thus its estimation at this initialization step. And as previously, the iterative procedure does not correct this effect.
 
 The results for option (c) (not shown here) are very similar to those obtained with our initialization procedure. This alternative is equivalent to include a constant term in the linear regression to estimate $f$. Its estimation is less degraded compared to option (b) and it is correct in the loop.  
 
 Our choice of estimating first the function $f$ using an unweighted regression is more flexible in the sense that it does not capture the all segmentation effect at the initialization step allowing thus the iterative procedure to correctly separate the function and the segmentation terms.\\
  
 \item[(3) Function model] The sensitivity of the procedure to the initialization step discussed above highlights the possible confusion between the function and segmentation. This sensitivity can be further explored by testing different models for $f$. The idea behind is that simpler models might be less confused with the segmentation making the procedure more accurate in terms of change-point locations. We tested two alternatives: (a) the shape of $f$ is known up to a scaling factor, i.e. $f_t=a_1\cos(2\pi t/L)$; (b) the statistically significant terms of the Fourier series are selected which have a p-value < $0.001$. Figure \ref{fig:cos_out_init} and \ref{fig:selb_out_init} show that the results for these two cases are both consistent and improve the segmentation results compared to our method (see Figure \ref{fig:quality_criteria} and  \ref{fig:RMSE_f}) as expected. Especially, the overall RMSE of the fitted function is strongly reduced. The impact on the positions and amplitudes of the change-points is rather small, however, and the impact in the case of real data is negligible (see Section \ref{sec:real}). This test points to the importance of the function model in our method. However, when it comes to real data, the real form of the function is not well known, i.e. the Fourier series of order 4 or even higher may be inadequate. It might thus be useful in a future version of the method to use a more complex base of functions. 
\end{description}

\begin{thebibliography}{Tototo}

\bibitem[Ardia {\em et~al.} (2019)]{ardia2019frequentist}%
{\sc Ardia, D.}, {\sc Dufays, A.} and {\sc Criado, C.~O.}
\newblock (2019).
\newblock Frequentist and bayesian change-point models: A missing link.
\newblock

\bibitem[Arlot and Massart (2009)]{Arlot2009}%
{\sc Arlot, S.} and {\sc Massart, P.}
\newblock (2009).
\newblock Data-driven calibration of penalties for least-squares regression.
\newblock {\em J. Mach. Learn. Res.}
\newblock {\bf 10} 245--279.

\bibitem[Auger and Lawrence (1989)]{Auger1989}%
{\sc Auger, I.~E.} and {\sc Lawrence, C.~E.}
\newblock (Jan, 1989).
\newblock Algorithms for the optimal identification of segment neighborhoods.
\newblock {\em Bulletin of Mathematical Biology}.
\newblock {\bf 51}~{\bf (1)} 39--54.

\bibitem[Bai and Perron (2003)]{BaiPerron2003}%
{\sc Bai, J.} and {\sc Perron, P.}
\newblock (2003).
\newblock Computation and analysis of multiple structural change models.
\newblock {\em Journal of Applied Econometrics}.
\newblock {\bf 18}~{\bf (1)} 1--22.

\bibitem[Bertin {\em et~al.} (2017)]{Bertin2017}%
{\sc Bertin, K.}, {\sc Collilieux, X.}, {\sc Lebarbier, E.} and {\sc Meza, C.}
\newblock (2017).
\newblock Semi-parametric segmentation of multiple series using a dp-lasso
  strategy.
\newblock {\em Journal of Statistical Computation and Simulation}.
\newblock {\bf 87}~{\bf (6)} 1255--1268.

\bibitem[Bevis {\em et~al.} (1992)]{Bevis1992}%
{\sc Bevis, M.}, {\sc Businger, S.}, {\sc Herring, T.~A.}, {\sc Rocken, C.},
  {\sc Anthes, R.~A.} and {\sc Ware, R.~H.}
\newblock (1992).
\newblock Gps meteorology: Remote sensing of atmospheric water vapor using the
  global positioning system.
\newblock {\em JOURNAL OF GEOPHYSICAL RESEARCH}.
\newblock {\bf 97}~{\bf (15)} 787--801.

\bibitem[Birgé and Massart (2001)]{BM2001}%
{\sc Birgé, L.} and {\sc Massart, P.}
\newblock (2001).
\newblock Gaussian model selection.
\newblock {\em Journal of the European Mathematical Society}.
\newblock {\bf 3} 203--268.

\bibitem[Bock (2016)]{Bock2016}%
{\sc Bock, O.}, (2016).
\newblock Gps data: Daily and monthly reprocessed iwv data from 120 global gps
  stations, version 1.2.

\bibitem[Bock {\em et~al.} (2013)]{Bock2013}%
{\sc Bock, O.}, {\sc Bosser, P.}, {\sc Bourcy, T.}, {\sc David, L.}, {\sc
  Goutail, F.}, {\sc Hoareau, C.}, {\sc Keckhut, P.}, {\sc Legain, D.}, {\sc
  Pazmino, A.}, {\sc Pelon, J.}, {\sc Pipis, K.}, {\sc Poujol, G.}, {\sc
  Sarkissian, A.}, {\sc Thom, C.}, {\sc Tournois, G.} and {\sc Tzanos, D.}
\newblock (2013).
\newblock Accuracy assessment of water vapour measurements from in situ and
  remote sensing techniques during the demevap 2011 campaign at ohp.
\newblock {\em Atmospheric Measurement Techniques}.
\newblock {\bf 6}~{\bf (10)} 2777--2802.

\bibitem[Bock {\em et~al.} (2018)]{Bock2018}%
{\sc Bock, O.}, {\sc Collilieux, X.}, {\sc Guillamon, F.}, {\sc Lebarbier, E.}
  and {\sc Pascal, C.}
\newblock (2018).
\newblock A breakpoint detection in the mean model with heterogeneous variance
  on fixed time-intervals.
\newblock {\em Statistics and Computing}.
\newblock {\bf 63}~{\bf (1)} 22--32.

\bibitem[Bock and Parracho (2019)]{Bock2019}%
{\sc Bock, O.} and {\sc Parracho, A.}
\newblock (2019).
\newblock Consistency and representativeness of integrated water vapour from
  ground-based gps observations and era-interim reanalysis.
\newblock {\em Atmos. Chem. Phys.}
\newblock {\bf 19} 9453–9468.

\bibitem[Caussinus and Mestre (2004)]{CaussinusMestre2004}%
{\sc Caussinus, H.} and {\sc Mestre, O.}
\newblock (2004).
\newblock Detection and correction of artificial shifts in climate series.
\newblock {\em Journal of the Royal Statistical Society: Series C (Applied
  Statistics)}.
\newblock {\bf 53}~{\bf (3)} 405--425.

\bibitem[Chakar {\em et~al.} (2017)]{Chakar2017}%
{\sc Chakar, S.}, {\sc Lebarbier, E.}, {\sc L{\'e}vy-Leduc, C.}, {\sc Robin,
  S.} {\em et~al.}
\newblock (2017).
\newblock A robust approach for estimating change-points in the mean of an
  {A}{R}(1) process.
\newblock {\em Bernoulli}.
\newblock {\bf 23}~{\bf (2)} 1408--1447.

\bibitem[Costa and Soares (2009)]{Costa2009}%
{\sc Costa, A.~C.} and {\sc Soares, A.}
\newblock (Apr, 2009).
\newblock Homogenization of climate data: Review and new perspectives using
  geostatistics.
\newblock {\em Mathematical Geosciences}.
\newblock {\bf 41}~{\bf (3)} 291--305.

\bibitem[Dee {\em et~al.} (2011)]{Dee2011}%
{\sc Dee, D.~P.}, {\sc Uppala, S.}, {\sc Simmons, A.}, {\sc Berrisford, P.},
  {\sc Poli, P.}, {\sc Kobayashi, S.}, {\sc Andrae, U.}, {\sc Balmaseda, M.},
  {\sc Balsamo, G.}, {\sc Bauer, d.~P.} {\em et~al.}
\newblock (2011).
\newblock The era-interim reanalysis: Configuration and performance of the data
  assimilation system.
\newblock {\em Quarterly Journal of the royal meteorological society}.
\newblock {\bf 137}~{\bf (656)} 553--597.

\bibitem[Easterling and Peterson (1995)]{Easterling1995}%
{\sc Easterling, D.} and {\sc Peterson, T.}
\newblock (1995).
\newblock A new method for detecting undocumented discontinuities in
  climatological time series.
\newblock {\em Int. J. Climatol.}
\newblock {\bf 15} 369--377.

\bibitem[Estey and Meertens (1999)]{esteymertens1999}%
{\sc Estey, L.} and {\sc Meertens, C.}
\newblock (1999).
\newblock Teqc: the multi-purpose toolkit for gps/glonass data.
\newblock {\em GPS Solutions}.
\newblock {\bf 3} 42--49.

\bibitem[Gazeaux {\em et~al.} (2015)]{gazeaux2015joint}%
{\sc Gazeaux, J.}, {\sc Lebarbier, E.}, {\sc Collilieux, X.} and {\sc
  M{\'e}tivier, L.}
\newblock (2015).
\newblock Joint segmentation of multiple gps coordinate series.
\newblock {\em Journal de la Soci{\'e}t{\'e} Fran{\c{c}}aise de Statistique}.
\newblock {\bf 156}~{\bf (4)} 163--179.

\bibitem[Jones {\em et~al.} (1986)]{Jones1986}%
{\sc Jones, P.~D.}, {\sc Raper, S. C.~B.}, {\sc Bradley, R.~S.}, {\sc Diaz,
  H.~F.}, {\sc Kellyo, P.~M.} and {\sc Wigley, T. M.~L.}
\newblock (1986).
\newblock Northern hemisphere surface air temperature variations: 1851–1984.
\newblock {\em Journal of Climate and Applied Meteorology}.
\newblock {\bf 25}~{\bf (2)} 161--179.

\bibitem[Killick {\em et~al.} (2012)]{Killick2012}%
{\sc Killick, R.}, {\sc Fearnhead, P.} and {\sc Eckley, I.~A.}
\newblock (2012).
\newblock Optimal detection of changepoints with a linear computational cost.
\newblock {\em Journal of the American Statistical Association}.
\newblock {\bf 107}~{\bf (500)} 1590--1598.

\bibitem[Lavielle (2005)]{Lavielle2005}%
{\sc Lavielle, M.}
\newblock (2005).
\newblock Using penalized contrasts for the change-point problem.
\newblock {\em Signal Processing}.
\newblock {\bf 85}~{\bf (8)} 1501--1510.

\bibitem[Lebarbier (2005)]{Lebarbier2005}%
{\sc Lebarbier, E.}
\newblock (2005).
\newblock Detecting multiple change-points in the mean of {G}aussian process by
  model selection.
\newblock {\em Signal Processing}.
\newblock {\bf 85} 717--736.

\bibitem[Li and Lund (2012)]{LiLund2012}%
{\sc Li, S.} and {\sc Lund, R.}
\newblock (2012).
\newblock Multiple changepoint detection via genetic algorithms.
\newblock {\em Journal of Climate}.
\newblock {\bf 25}~{\bf (2)} 674--686.

\bibitem[Lu {\em et~al.} (2010)]{Lu2010}%
{\sc Lu, Q.}, {\sc Lund, R.} and {\sc Lee, T. C.~M.}
\newblock (2010).
\newblock An mdl approach to the climate segmentation problem.
\newblock {\em The Annals of Applied Statistics}.
\newblock {\bf 4}~{\bf (1)} 299--319.

\bibitem[Maidstone {\em et~al.} (2017)]{Maidstone2017}%
{\sc Maidstone, R.}, {\sc Hocking, T.}, {\sc Rigaill, G.} and {\sc Fearnhead,
  P.}
\newblock (2017).
\newblock On optimal multiple changepoint algorithms for large data.
\newblock {\em Stat. Comput.}
\newblock {\bf 27} 519–533.

\bibitem[Menne and Williams (2005)]{Menne2005}%
{\sc Menne, M.~J.} and {\sc Williams, C.~N.}
\newblock (2005).
\newblock Detection of undocumented changepoints using multiple test statistics
  and composite reference series.
\newblock {\em Journal of Climate}.
\newblock {\bf 18}~{\bf (20)} 4271--4286.

\bibitem[Ning {\em et~al.} (2016a)]{ning2016}%
{\sc Ning, T.}, {\sc Wickert, J.}, {\sc Deng, Z.}, {\sc Heise, S.}, {\sc Dick,
  G.}, {\sc Vey, S.} and {\sc Schöne, T.}
\newblock (2016a).
\newblock Homogenized time series of the atmospheric water vapor content
  obtained from the gnss reprocessed data.
\newblock {\em Journal of Climate}.
\newblock {\bf 29}~{\bf (7)} 2443--2456.

\bibitem[Ning {\em et~al.} (2016b)]{NingUncertainty2016}%
{\sc Ning, T.}, {\sc Wang, J.}, {\sc Elgered, G.}, {\sc Dick, G.}, {\sc
  Wickert, J.}, {\sc Bradke, M.}, {\sc Sommer, M.}, {\sc Querel, R.} and {\sc
  Smale, D.}
\newblock (2016b).
\newblock The uncertainty of the atmospheric integrated water vapour estimated
  from gnss observations.
\newblock {\em Atmos. Meas. Tech.}
\newblock {\bf 9}~{\bf (1)} 79--92.

\bibitem[Parracho {\em et~al.} (2018)]{parracho2018global}%
{\sc Parracho, A.~C.}, {\sc Bock, O.} and {\sc Bastin, S.}
\newblock (2018).
\newblock Global iwv trends and variability in atmospheric reanalyses and gps
  observations.
\newblock {\em Atmospheric Chemistry and Physics}.
\newblock {\bf 18}~{\bf (22)} 16213--16237.

\bibitem[Peterson {\em et~al.} (1998)]{Peterson1998}%
{\sc Peterson, T.~C.}, {\sc Easterling, D.~R.}, {\sc Karl, T.~R.}, {\sc
  Groisman, P.}, {\sc Nicholls, N.}, {\sc Plummer, N.}, {\sc Torok, S.}, {\sc
  Auer, I.}, {\sc Boehm, R.}, {\sc Gullett, D.} {\em et~al.}
\newblock (1998).
\newblock Homogeneity adjustments of in situ atmospheric climate data: a
  review.
\newblock {\em International Journal of Climatology: A Journal of the Royal
  Meteorological Society}.
\newblock {\bf 18}~{\bf (13)} 1493--1517.

\bibitem[Picard {\em et~al.} (2005)]{Picard2005}%
{\sc Picard, F.}, {\sc Robin, S.}, {\sc Lavielle, M.}, {\sc Vaisse, C.} and
  {\sc Daudin, J.-J.}
\newblock (Feb, 2005).
\newblock A statistical approach for array cgh data analysis.
\newblock {\em BMC Bioinformatics}.
\newblock {\bf 6}~{\bf (1)}~27.

\bibitem[Reeves {\em et~al.} (2007)]{Reeves2007}%
{\sc Reeves, J.}, {\sc Chen, J.}, {\sc Wang, X.~L.}, {\sc Lund, R.} and {\sc
  Lu, Q.~Q.}
\newblock (2007).
\newblock A review and comparison of changepoint detection techniques for
  climate data.
\newblock {\em Journal of Applied Meteorology and Climatology}.
\newblock {\bf 46}~{\bf (6)} 900--915.

\bibitem[Rigaill (2015)]{Rigaill2015}%
{\sc Rigaill, G.}
\newblock (2015).
\newblock A pruned dynamic programming algorithm to recover the best
  segmentations with $1$ to $k_max$ change-points.
\newblock {\em Journal de la Société Française de Statistique}.
\newblock {\bf 156}~{\bf (4)} 180--205.

\bibitem[Rissanen (1978)]{Rissanen78}%
{\sc Rissanen, J.}
\newblock (1978).
\newblock Modelling by the shortest data description.
\newblock {\em Automatica}.
\newblock {\bf 14} 465–471.

\bibitem[Rousseeuw and Croux (1993)]{Rousseeuw1993}%
{\sc Rousseeuw, P.~J.} and {\sc Croux, C.}
\newblock (1993).
\newblock Alternatives to the median absolute deviation.
\newblock {\em Journal of the American Statistical Association}.
\newblock {\bf 88}~{\bf (424)} 1273--1283.

\bibitem[Szentimrey (2008)]{Szentimrey2008}%
{\sc Szentimrey, T.}
\newblock (2008).
\newblock Development of mash homogenization procedure for daily data.
  proceedings of the fifth seminar for homogenization and quality control in
  climatological databases.
\newblock {\em WCDMP-No. 71}.
\newblock  123–130.

\bibitem[Truong {\em et~al.} (2020)]{Truong2019}%
{\sc Truong, C.}, {\sc Oudre, L.} and {\sc Vayatis, N.}
\newblock (2020).
\newblock Selective review of offline change point detection methods.
\newblock {\em Signal Processing}.
\newblock {\bf 167} 107299.

\bibitem[Van~Malderen {\em et~al.} (2014)]{VanMalderen2014}%
{\sc Van~Malderen, R.}, {\sc Brenot, H.}, {\sc Pottiaux, E.}, {\sc Beirle, S.},
  {\sc Hermans, C.}, {\sc De~Maziere, M.}, {\sc Wagner, T.}, {\sc De~Backer,
  H.} and {\sc Bruyninx, C.}
\newblock (08, 2014).
\newblock A multi-site intercomparison of integrated water vapour observations
  for climate change analysis.
\newblock {\bf 7}.

\bibitem[Varadhan and Roland (2008)]{varadhan2008simple}%
{\sc Varadhan, R.} and {\sc Roland, C.}
\newblock (2008).
\newblock Simple and globally convergent methods for accelerating the
  convergence of any em algorithm.
\newblock {\em Scandinavian Journal of Statistics}.
\newblock {\bf 35}~{\bf (2)} 335--353.

\bibitem[Venema {\em et~al.} (2012)]{Venema2012}%
{\sc Venema, V. K.~C.}, {\sc Mestre, O.}, {\sc Aguilar, E.}, {\sc Auer, I.},
  {\sc Guijarro, J.~A.}, {\sc Domonkos, P.}, {\sc Vertacnik, G.}, {\sc
  Szentimrey, T.}, {\sc Stepanek, P.}, {\sc Zahradnicek, P.}, {\sc Viarre, J.},
  {\sc M\"uller-Westermeier, G.}, {\sc Lakatos, M.}, {\sc Williams, C.~N.},
  {\sc Menne, M.~J.}, {\sc Lindau, R.}, {\sc Rasol, D.}, {\sc Rustemeier, E.},
  {\sc Kolokythas, K.}, {\sc Marinova, T.}, {\sc Andresen, L.}, {\sc Acquaotta,
  F.}, {\sc Fratianni, S.}, {\sc Cheval, S.}, {\sc Klancar, M.}, {\sc Brunetti,
  M.}, {\sc Gruber, C.}, {\sc Prohom~Duran, M.}, {\sc Likso, T.}, {\sc Esteban,
  P.} and {\sc Brandsma, T.}
\newblock (2012).
\newblock Benchmarking homogenization algorithms for monthly data.
\newblock {\em Climate of the Past}.
\newblock {\bf 8}~{\bf (1)} 89--115.

\bibitem[Vey {\em et~al.} (2009)]{Vey2009}%
{\sc Vey, S.}, {\sc Dietrich, R.}, {\sc Fritsche, M.}, {\sc Rülke, A.}, {\sc
  Steigenberger, P.} and {\sc Rothacher, M.}
\newblock (2009).
\newblock On the homogeneity and interpretation of precipitable water time
  series derived from global gps observations.
\newblock {\bf 114}~{\bf (D10)}.

\bibitem[Wang (2008)]{Wang2008}%
{\sc Wang, X.~L.}
\newblock (2008).
\newblock Accounting for autocorrelation in detecting mean shifts in climate
  data series using the penalized maximal t or f test.
\newblock {\em Journal of Applied Meteorology and Climatology}.
\newblock {\bf 47}~{\bf (9)} 2423--2444.

\bibitem[Weatherhead {\em et~al.} (1998)]{Weatherhead1998}%
{\sc Weatherhead, E.~C.}, {\sc Reinsel, G.~C.}, {\sc Tiao, G.~C.}, {\sc Meng,
  X.}, {\sc Choi, D.}, {\sc Cheang, W.}, {\sc Keller, T.}, {\sc DeLuisi, J.},
  {\sc Wuebbles, D.~J.}, {\sc Kerr, J.~B.}, {\sc Miller, A.~J.}, {\sc Oltmans,
  S.~J.} and {\sc Frederick, J.~E.}
\newblock (1998).
\newblock Factors affecting the detection of trends: Statistical considerations
  and applications to environmental data.
\newblock

\bibitem[Zhang and Siegmund (2007)]{Zhang2007}%
{\sc Zhang, N.~R.} and {\sc Siegmund, D.~O.}
\newblock (2007).
\newblock A modified {B}ayes information criterion with applications to the
  analysis of comparative genomic hybridization data.
\newblock {\em Biometrics}.
\newblock {\bf 63}~{\bf (1)} 22--32.

\end{thebibliography}
\end{document}